\documentclass[acmsmall]{acmart}

\AtBeginDocument{%
  }

\setcopyright{acmlicensed}
\copyrightyear{2024}
\acmYear{2024}
\acmDOI{XXXXXXX.XXXXXXX}

\acmJournal{JACM}
\acmVolume{37}
\acmNumber{4}
\acmArticle{111}
\acmMonth{8}

\usepackage{booktabs}
\usepackage{threeparttable}
\usepackage{pifont}

\begin{document}
\title{AI Ethics in Smart Homes: Progress, User Requirements and Challenges}

\author{Liqian You}
\affiliation{%
  \institution{University of Technology, Sydney}
  \city{Sydney}
  \country{Australia}}
\email{Liqian.You@student.uts.edu.au}

\author{Jianlong Zhou}
\affiliation{%
  \institution{University of Technology, Sydney}
  \city{Sydney}
  \country{Australia}}
\email{Jianlong.Zhou@uts.edu.au}

\author{Zhiwei Li}
\affiliation{%
  \institution{University of Technology, Sydney}
  \city{Sydney}
  \country{Australia}}
\email{Zhiwei.Li@student.uts.edu.au}

\author{Fang Chen}
\affiliation{%
  \institution{University of Technology, Sydney}
  \city{Sydney}
  \country{Australia}}
\email{Fang.Chen@uts.edu.au}

\begin{abstract}
With the rise of Internet of Things (IoT) technologies in smart homes and the integration of artificial intelligence (AI), ethical concerns have become increasingly significant. This paper explores the ethical implications of AI-driven detection technologies in smart homes using the User Requirements Notation (URN) framework. In this paper, we thoroughly conduct thousands of related works from 1985 to 2024 to identify key trends in AI ethics, algorithm methods, and technological advancements. The study presents an overview of smart home and AI ethics, comparing traditional and AI-specific ethical issues, and provides guidelines for ethical design across areas like privacy, fairness, transparency, accountability, and user autonomy, offering insights for developers and researchers in smart homes.
\end{abstract}



\received{20 February 2007}
\received[revised]{12 March 2009}
\received[accepted]{5 June 2009}

\maketitle

\section{Introduction}

Over the past several years, the term "SMART" is not merely a household system with some automatic functions \cite{sun_systematic_2021}, but has become homogeneous with any innovative Artificial Intelligence-driven (AI-driven) technology or system \cite{agarwal2021edge} as well. The capability of collecting information and data from home surroundings and interacting correspondingly is the crucial feature of smart technology \cite{chakraborty_smart_2023}. Smart homes refer to habitations or residences equipped with automatic technological devices that can meet people’s requirements and improve their daily living in different aspects \cite{chakraborty_smart_2023}. Smart home systems consist of various electrical devices, including lights, heating, air conditioners, security systems, entertainment systems, and kitchen appliances \cite{hasan_low-cost_2019}. Basically, security systems include cameras, doors, and lockers, while entertainment systems include televisions, audio systems, and projectors \cite{hasan_smart_2019}. These systems allow residents to control, operate, and monitor devices from a distance, ensuring convenience, well-being, comfort, efficiency, security, energy-saving, and high quality of life \cite{nimmy_lightweight_2022}. Smart home technologies can interact with one another, and are managed through networks such as Bluetooth, Wi-Fi, Zigbee, etc. \cite{sun_systematic_2021}. Furthermore, smart homes generally integrate various sensors, along with AI algorithms \cite{hunt2014artificial} to automatically control electronic devices and electrical appliances, learning to communicate users' personalized data through their preferences and improve using experiences \cite{sun_systematic_2021}. Building on this foundation, AI detection systems \cite{elrawy2018intrusion}, including image detection \cite{chauhan2018convolutional} related to smart home healthcare \cite{amiribesheli_review_2015}, face detection in home security systems \cite{wati2017design}, and voice detection in home animation systems \cite{malik2020light} have been widely utilized in modern smart homes in recent years. Other detection systems, including gesture detection, movement obstacle detection, and live streaming remote control, are rapidly penetrating people’s daily lives as well \cite{hasan_smart_2019}. These AI-based detection systems in smart homes have drawn a leading topic in the field of the Internet of Things (IoTs) \cite{madakam2015internet}.

Increasing demand for utilizing the IoTs in smart homes has raised ethical challenges in developing detection technologies over recent decades \cite{chan2008review}. This paper follows the User Requirements Notation (URN) \cite{amyot2011user}framework to explore user concerns and ethical implications, analyzing AI ethics through both user-centric and technological perspectives. By examining the relationship between AI-driven technologies and ethics, this paper identifies potential risks and challenges, proposing actionable guidelines and principles for designing human-centered smart home detection systems. The discussion delves into the interplay between general ethics and AI-specific ethics across five dimensions: decision-making, accountability, bias and fairness, transparency, and unforeseen risks. We also present ethical design guidelines, emphasizing core principles such as Privacy by Design, Fairness in Algorithmic Decision-Making, Transparency and Explainability, Accountability, and Human Autonomy. The paper provides a comprehensive analysis of AI ethics in the context of smart homes, IoT technologies, detection systems, healthcare, and AI innovations, integrating URN to ensure responsible development. It concludes by highlighting key contributions and proposing future research directions to safeguard privacy, ensure fairness, and foster trust, aiming to establish a foundation for ethically sound smart home technologies.

To outline, this paper makes contributions to:
\begin{itemize}
    \item Analyze the evolution of AI ethics in smart homes by reviewing peer-reviewed articles, providing an extensive literature foundation for future research.
    \item Establish a comprehensive foundation for AI ethics research, particularly in smart home detection technologies.
    \item Offer heuristic and empirical guidance for scholars and developers exploring AI-driven detection systems in smart homes.
    \item Present a framework for integrating ethical considerations, such as privacy, autonomy, transparency, bias, and fairness, into AI-based smart home technologies.
    \item Propose ethical design guidelines for AI detection systems, emphasizing principles such as Privacy by Design, Fairness in Algorithmic Decision-Making, Transparency and Explainability, Accountability, and Human Autonomy.
    \item Identify key challenges and risks in the development and deployment of AI technologies in smart homes, offering practical insights for responsible design.
\end{itemize}

\section{Background}

\subsection{Smart home technologies}

Kevin Ashton described IoTs as a network connecting physical objects, or "things," equipped with sensors and software, enabling them to collect and exchange data over the internet \cite{li2015internet}. In 2009, Ashton predicted that IoTs would revolutionize the world, much like the Internet had done, a transformation that is already unfolding \cite{sharma2019history}.

Lynn et al. \cite{lynn_cloud_thing_2020} reviewed the definition of IoTs in two layers: (1) the technical layer that IoTs is considered a combination of technical artefacts \cite{weyrich_reference_2016}; (2) the socio-technical layer that IoTs refer to leaders as the bit part of communication data between objects and the network \cite{shin_socio-technical_2014}. Nevertheless, the concept of IoTs includes both the approach of things that are capable of being connected to the network and the approach of usability and functionality which can meet users' requirements with objects connected to the network \cite{ibarra-esquer_tracking_2017}. The objects related to IoTs that can be linked to the network including large machines, electrical devices, mobile applications, personal computers, and virtual or physical objects with different sensors that can exchange data through the Internet and have the capability of creating, requesting, consuming, redirecting, or accessing digital information \cite{elkhodr_internet_2013}. Moreover, other literature presents that IoTs stand for "smart", for instance, smart homes \cite{alam_review_2012}, smart cities \cite{yin2015literature}, and smart campuses \cite{muhamad2017smart}. "Smart" is a word that refers to things or objects calculating, connecting, and communicating with one another over the network \cite{chakraborty_smart_2023}. These smart AI-driven technologies or systems, also regarded as the solution of IoTs, have been widely utilized in our daily lives in recent years.

The \textit{Household Penetration Rate by Segment} \cite{statista_research_department_global_2023} in the worldwide smart home market is predicted to increase between 2024 and 2028 by a total of 14.31\%. The penetration rate of the smart homes market in 2024 is 18.89\%, and it is expected to reach 33.2\% in 2028 \cite{statista_research_department_global_2023-1}. The global revenue in the smart home market is expected to increase from 2023 to 2028 by 96.7 billion U.S. dollars (increasing 71.74\%), and it will reach 231.6 billion U.S. dollars in 2028 \cite{statista_research_department_global_2023-1}. The number of users in the smart home market was 360.66 million in 2023, and it is projected to grow to 785.16 million by 2028, representing an increase of 117.69\% over this five-year period \cite{statista_research_department_global_2023-2}. This significant growth highlights the increasing demand for smart home technologies, driven by factors such as advancements in IoTs, AI \cite{hunt2014artificial}, and home automation \cite{aldrich2003smart}.

A comprehensive range of IoT solutions in smart homes encompasses hardware components (e.g., processors, sensors, gateways, controllers, switches), software platforms (independent or centralized with front-end and back-end interfaces), and services such as automated device management, remote control and monitoring, and iterative system deployment and updates \cite{bucharest_university_of_economic_studies_romania_adopting_2021}. These IoTs bring about high efficiency and productivity in daily lives, improving our physical and mental health, maintaining more convenient transportation, saving energy and making our lives more comfortable \cite{alam_review_2012}. Researchers have proposed automation in smart homes since the 80's \cite{jacobsson2016risk, pirbhulal2016novel, jabbar2019design}. Meanwhile, the solutions of IoTs in smart homes have drawn attention in recent decades. Smart homes can be defined as the iteration of traditional household automation, while AI-driven IoTs induce the further development of integrated control and management \cite{horrigan1987home}. Smart homes are defined as residences with a variety of automatic technological devices \cite{aldrich2003smart} to control the house in the aspects of lights, temperature, humidity, security cameras, entertainment systems, kitchen appliances, etc.

As the number of IoT devices in smart homes continues to grow rapidly, and smart home-based detection technologies increasingly become part of everyday life, it is essential for researchers and scholars to explore the associated AI ethical issues in greater depth. While numerous papers on AI-driven detection technologies in smart homes primarily focus on technical aspects, such as architecture \cite{ahanger_iot-inspired_2020}, frameworks \cite{strengers2013smart}, functionalities \cite{sowah2020design}, and advanced algorithms \cite{ahanger_iot-inspired_2020}, there is a notable lack of discussion surrounding the ethical considerations, particularly within the realm of detection technologies. While some regional statistics \cite{statista_research_department_global_2023-1, statista_research_department_global_2023, statista_research_department_global_2023-2} and market reports\cite{mordor2020consumer, shin2018will} address the adoption, acceptance, and usability of smart home IoTs, the number of scientific studies delving into the ethical dimensions remains limited, often concentrating on specific user groups such as younger generations \cite{mkacik2017adoption}, elderly individuals with disabilities, or patients \cite{bucharest_university_of_economic_studies_romania_adopting_2021}. 

To address this gap, this paper conducts a broader investigation by reviewing a diverse range of publications that cover various demographic groups and backgrounds. Through this analysis, we aim to provide a more comprehensive discussion of the ethical considerations of AI-driven detection technologies in smart homes, with an emphasis on ensuring that these technologies are developed and deployed responsibly.

\subsection{Ethics and AI ethics}

\subsubsection{General Ethics} Ethics, traditionally rooted in philosophy \cite{shafer2006ethics}, addresses fundamental questions about morality, such as determining what is right or wrong, good or bad, and the responsibilities individuals owe to one another \cite{deigh2010introduction}. It entails a systematic and critical examination of moral dilemmas encountered in everyday life and within professional fields like business, medicine, and technology \cite{reynolds2006moral}. This discipline provides a robust framework for evaluating actions not only in terms of their effectiveness or legality but also through the lens of moral responsibility, thus guiding human conduct, shaping public policies, and setting standards for professional behavior \cite{deigh2010introduction}. By integrating ethical principles, societies aim to promote fairness, accountability, and the well-being of individuals and communities.

Ethics can be categorized into three main areas: normative ethics, meta-ethics, and applied ethics \cite{bonhoeffer1995ethics}. Normative ethics focuses on determining what actions are morally right or wrong by establishing standards or norms for behavior \cite{posner1997standards}. It addresses practical questions, such as what one ought to do in particular situations, drawing from moral theories like utilitarianism (maximizing overall happiness), deontology (emphasizing duties and rules), and virtue ethics (highlighting moral character) \cite{kagan2018normative}. Meta-ethics, on the other hand, explores the nature and origins of ethical principles, delving into questions about the meaning of moral statements, the existence of moral truths, and how such truths can be known independently of individual opinions \cite{o2001meta}. Applied ethics deals with the application of ethical theory to specific controversial issues, such as medical ethics, business ethics, and environmental ethics, offering guidance on moral challenges faced by individuals or societies by leveraging insights from normative ethics \cite{singer1986applied}.

General ethics has long played a vital role in shaping human behavior and decision-making by providing a moral framework for evaluating actions \cite{blasi1983moral}, rather than relying solely on considerations of effectiveness, legality, or customary practices \cite{deigh2010introduction}. Foundational documents like the Belmont Report \cite{paxton2020belmont} and the Declaration of Helsinki \cite{goodyear2007declaration} have guided ethical research practices for decades. However, the advent of big data and advanced technological methods has introduced new ethical complexities, particularly in fields like behavioral research \cite{cozby2012methods}. These developments have made it increasingly difficult to safeguard the rights and welfare of research participants, raising the need for renewed attention to ethical considerations in modern contexts \cite{favaretto2020first}.

\subsubsection{AI ethics} AI has impacted fields such as facial recognition \cite{kaur2020facial}, medical diagnostics \cite{kononenko2001machine}, and autonomous driving \cite{yurtsever2020survey}, contributing to improvements in economic, social, and safety aspects of human life. However, alongside these advancements, AI faces challenges related to explainability, data biases, and privacy concerns \cite{arrieta2020explainable}. AI ethics, which is an emerging direction \cite{hagendorff2020ethics}, seeks to address these issues by developing specific ethical frameworks to ensure AI systems operate within morally sound boundaries \cite{miller2001boundaries}. The goal is to identify and mitigate ethical concerns, ensuring that AI technologies adhere to established ethical standards throughout their lifecycle \cite{siau2020artificial}.

Favaretto et al. \cite{favaretto2020first} highlight the ethical challenges posed by Big Data research, emphasizing the need for updated guidelines that reflect the complexities of modern digital research. Traditional ethical principles must evolve to address the unique challenges introduced by technologies that distance researchers from their subjects, such as those found in AI and Big Data \cite{favaretto2020first}.

Similarly, Hagendorff et al. \cite{hagendorff2020ethics} provide a semi-systematic evaluation of 22 AI ethics guidelines, developed in response to the rapid advancements in AI systems. He emphasizes the pivotal role of ethical principles in shaping the research, development, and deployment of AI technologies, emphasizing the necessity of translating these guidelines into actionable and practical solutions to ensure their effectiveness in real-world applications.

\begin{table*}
\centering
\scriptsize
\caption{Comparison of general ethics and ethics in AI-driven technologies.}
\label{tab:comparison}
\begin{tabular}{p{1.4cm}p{4.5cm}p{6.9cm}} 
\hline
\textbf{Aspect} & \textbf{General Ethics} & \textbf{Ethics in AI-driven Technologies} \\ 
\hline
\textbf{Decision-Making Process}\cite{duan2019artificial}& Human reasoning; Empathy; Moral judgment; Philosophical principles; Cultural norms; Personal values& Design algorithms; Design systems capable of autonomous decisions; Support human decision-making; Define ethical guidelines; Implement ethical guidelines; Ensure fairness, transparency and accountability\\ 
\textbf{Accountability and Responsibility}\cite{doshi2017accountability}& Straightforward, resting primarily with individuals or organizations; Answer for actions; Face the consequences for unethical behavior& Complex in AI systems; Shared among developers, organizations, and AI systems; Comprehensive frameworks and regulations; Frameworks to clarify roles and responsibilities\\ 
\textbf{Bias and Fairness}\cite{mehrabi2021survey}& Uphold fairness; Avoid bias from personal, cultural, or societal influences; Biases  inadvertently influence decisions; Biases lead to ethical dilemmas& AI operates based on data contain biases; Biases perpetuated or amplified by AI algorithms; Result in unfair outcomes; Address issues requires meticulous attention to data selection\\ 
\textbf{Transparency and Explainability}\cite{ehsan2021expanding}& Explain the rationale behind decisions; Cornerstones of traditional ethical decision-making; Justify individual's actions and provide clear reasons for their ethical choices& AI systems perceived as "black boxes"; Complicates the demand for transparency and explainability; AI decisions are transparent; Explain the underlying algorithms in human-understandable terms; Maintain trust and accountability requires transparent AI decisions\\ 
\textbf{Unintended Consequences and Unforeseen Risks}\cite{cheatham2019confronting}& Consider the potential consequences and risks associated with actions; Strive to foresee and mitigate possible negative outcomes; Hard to predict all possible outcomes or unintended consequences& AI systems make decisions based on patterns and correlations; Difficult for humans to predict or understand; 
Ethical frameworks include comprehensive risk assessments and mitigation strategies; Address privacy concerns, biases, and broader societal impacts; Mitigation strategies  account for the societal impacts that arise from AI technologies\\ 
\hline
\end{tabular}
\end{table*}

As shown in Table ~\ref{tab:comparison}, there are clear distinctions between general ethics, which are grounded in human reasoning, empathy, and cultural norms, and AI ethics, which require the design of autonomous systems that ensure fairness, transparency, and accountability. This distinction is particularly significant in the context of smart home technologies, where AI systems handle sensitive tasks such as healthcare monitoring \cite{islam2020development}, decision-making \cite{duan2019artificial}, and automated assistance \cite{nirmala2022artificial}, often for vulnerable populations like older adults.

In smart homes, ethical AI \cite{zhou2020survey} development requires systems that not only align with human values but also address privacy concerns, avoid biases, and ensure accountability for unintended consequences \cite{shneiderman2020bridging}. These systems must respect personal autonomy, provide unbiased decision-making (e.g., in healthcare recommendations), and offer clear explanations for the actions taken by AI. Developing such frameworks is essential to safeguarding user rights while optimizing the efficiency and safety of smart home environments \cite{powers2020ethics, hagendorff2020ethics, zhou2023ai}.
 
\subsection{Research Methods and Search Strategies}

This study follows systematic literature review (SLR) guidelines \cite{kitchenham_guidelines_2007}, employing a structured approach to identify and analyze relevant research. The review protocol begins with defining broad search keywords "smart homes" and "IoTs," which are further refined by including fundamental segments and synonyms like "smart-home," "AI-driven home," "home automation," "AI ethics," "ethical consideration," and "detection technology" \cite{kitchenham_systematic_2009}. This comprehensive keyword strategy ensures the inclusion of diverse perspectives on smart home and AI-driven detection technologies. Papers were sourced exclusively from reputable databases, including ResearchGate, IEEE Xplore, and the ACM Digital Library, with a focus on English-language publications from January 1, 1985, to April 30, 2024, ensuring the relevance of the findings to this rapidly evolving domain.

Subsequently, a quality analysis was conducted to refine the dataset. Clear inclusion and exclusion criteria were applied to ensure relevance and academic rigor. The final stage involved summarizing and visualizing the selected literature, analyzing key aspects such as authorship, publication years, and research trends. Timeline evolution tools were utilized to trace shifts in research focus over time. These outputs provide critical insights into the development of smart home technologies and their associated ethical considerations, particularly in relation to user requirements notation. The findings serve as a foundational resource for researchers exploring the intersection of AI ethics and smart home innovations.

\section{Smart Home Technologies and AI Ethics}

\subsection{Overview of Research in Smart Homes} Understanding the trajectory of smart home research is essential for identifying key innovations and addressing emerging challenges in this rapidly evolving field. By analyzing patterns and relationships within the literature, we uncover the thematic and temporal progression of research areas, highlighting the technologies and concepts driving smart home advancements.

\begin{figure}[!ht]
\centering
\begin{minipage}{0.45\textwidth}
\centering
\includegraphics[width=\linewidth]{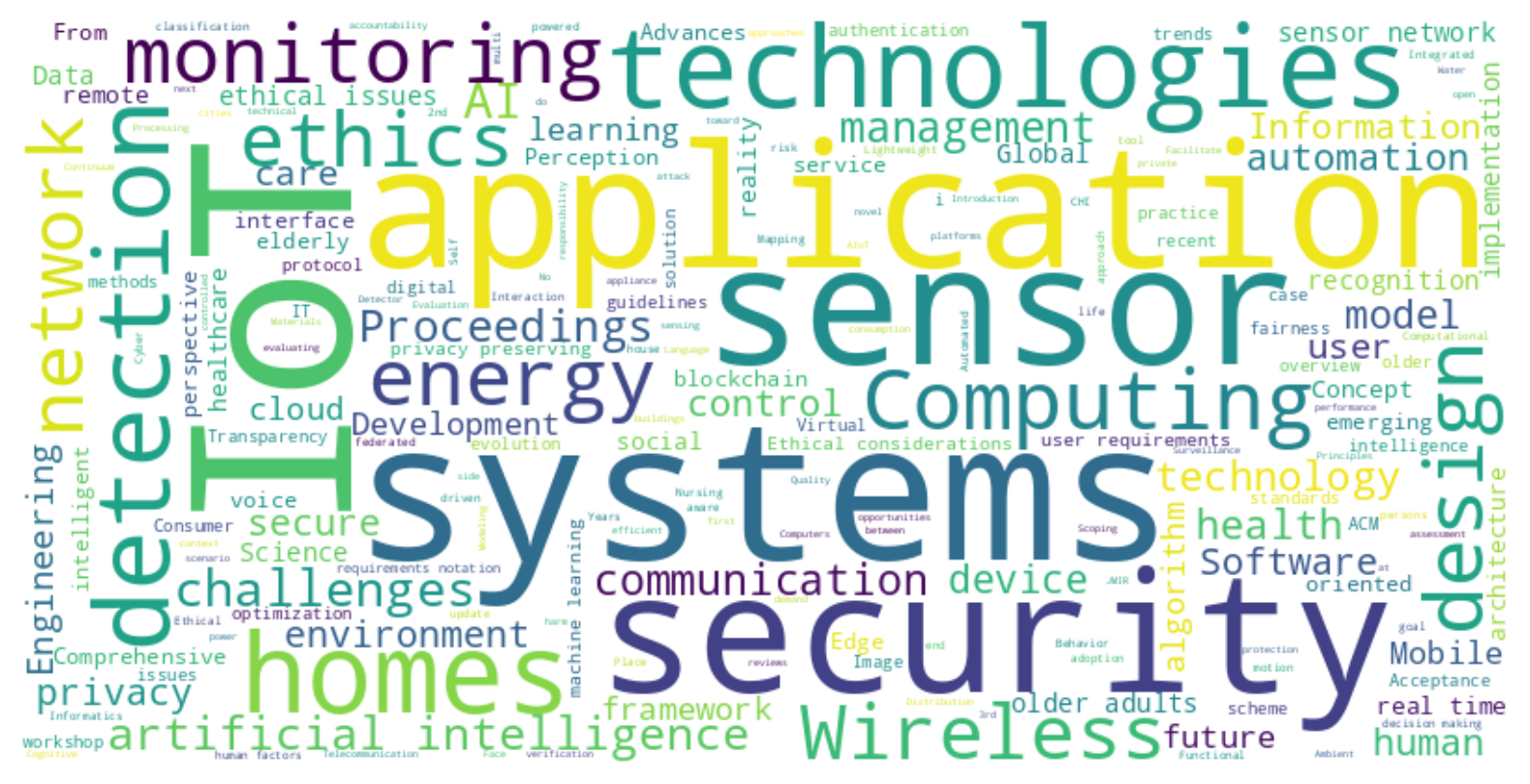}
\caption{\label{fig:word cloud}Word cloud of keywords from reviewed articles. With larger sizes indicating higher frequency, highlights the top keywords in smart home research.}
\Description{An image of keywords.}
\end{minipage}
\hfill
\begin{minipage}{0.5\textwidth}
\centering
\includegraphics[width=\linewidth]{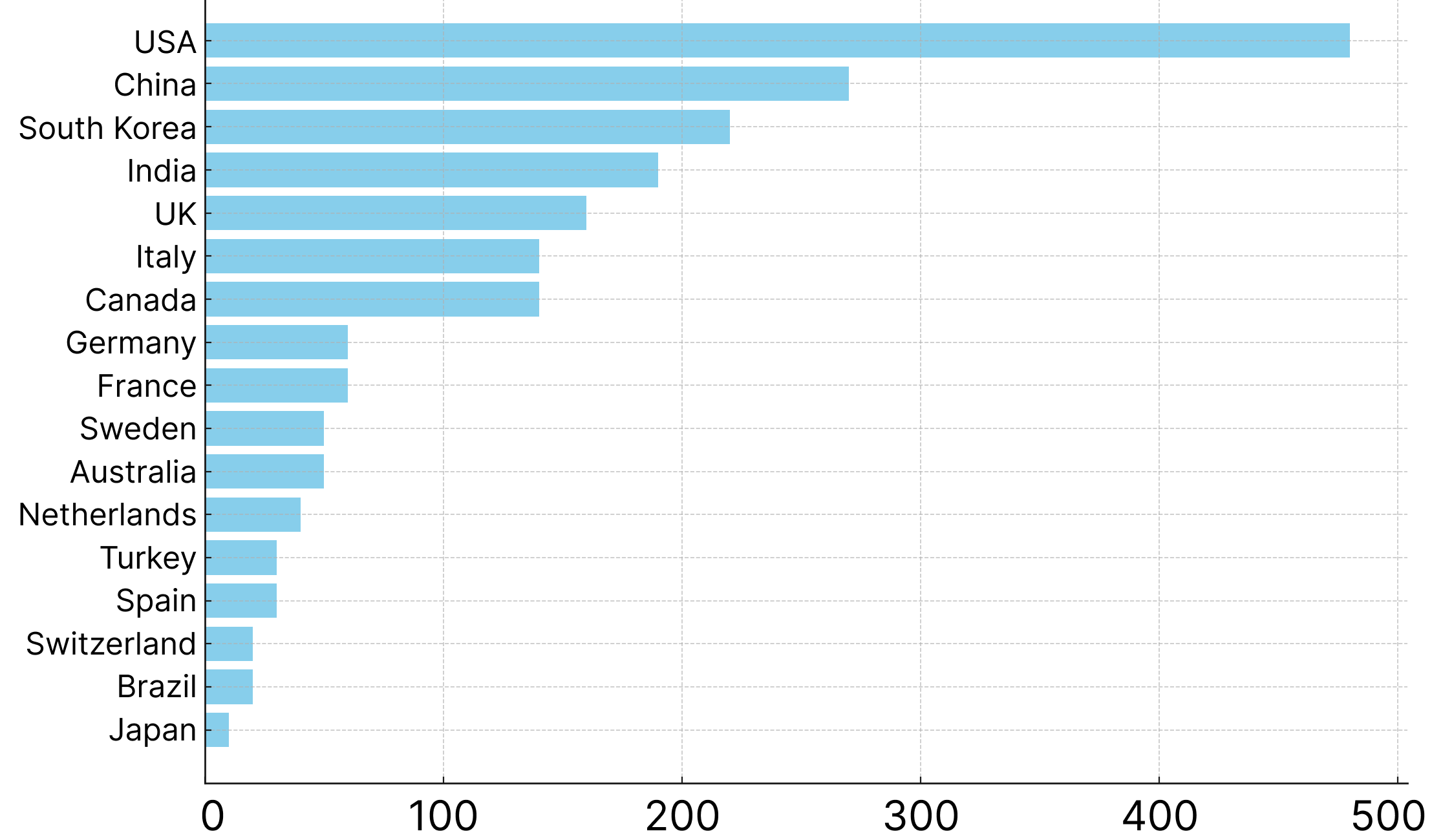}
\caption{\label{fig:country} Distribution of the number of publications from different countries.}
\Description{Chart of number of papers by country.}
\end{minipage}
\end{figure}

We initially collected 2289 papers and subsequently extracted 327 peer-reviewed articles from the datasets for detailed analysis, aiming to uncover trends, relationships, and themes in smart home research. To gain deeper insights, we visualized the data to explore connections between these articles. As illustrated in Figure ~\ref{fig:word cloud}, the word cloud showcases the most frequently occurring keywords in the reviewed literature, with larger text sizes representing higher frequencies. These keywords reveal the dominant research areas and themes within the field of smart homes, highlighting the critical role of interdisciplinary approaches that integrate advanced technologies, user-centric design, and ethical considerations.

According to Figure ~\ref{fig:country},  the top 5 countries based on the number of publications are the USA, China, South Korea, India and UK. These countries' vigorous technological infrastructures enable comprehensive research and development activities. 5G, WIFI and internet, widespread adoption of smart home appliances, and advanced telecommunications networks are essential for developing and evaluating smart home technologies. Besides, these countries' authorities and communities deliver considerable funding and policy support for research and development in novel technologies, particularly smart homes. For example, China's government has been prescient in moving forward with improvements in AI and IoTs \cite{merics2021connection}, both of which are elementary technologies that will contribute to the evolution of smart homes. 

\begin{figure}[!ht]
\centering
\includegraphics[width=0.9\linewidth]{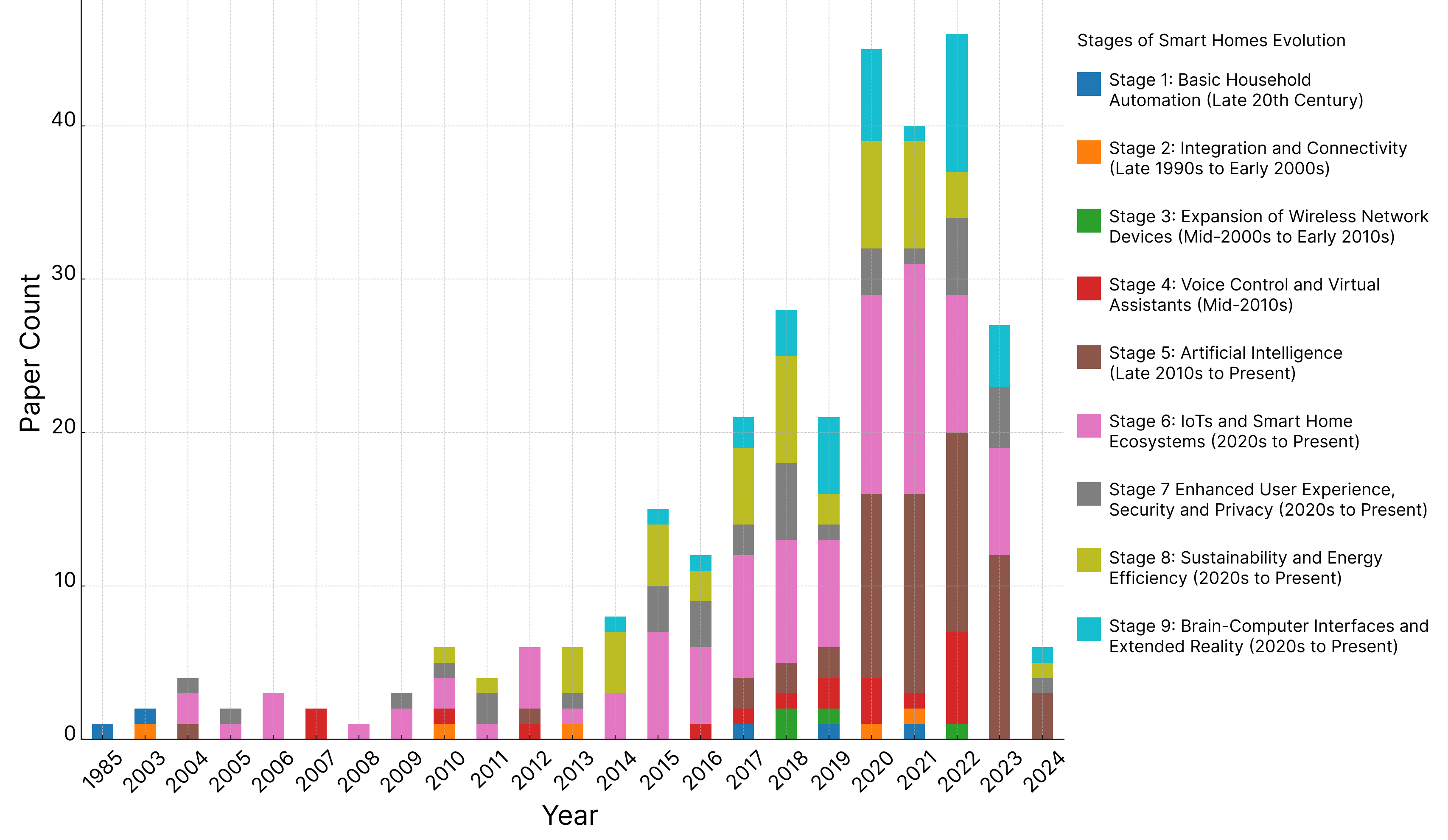}
\caption{\label{fig:pub} Annual distribution of peer-reviewed papers across nine stages of smart home evolution. With color-coded by stage highlighting yearly counts.}
\Description{}
\end{figure}

To capture the progression in smart home research, Figure ~\ref{fig:pub} presents a stacked bar chart illustrating annual publication trends across nine distinct stages of smart home evolution, as outlined in Figure ~\ref{fig:review evolution}. Each stage, representing a pivotal phase in the technological and conceptual advancement of smart homes, is color-coded for clarity, with the legend conveniently positioned on the right. The x-axis spans the years from 1985 to 2024, while the y-axis indicates the number of articles published. This visualization highlights the increasing research interest over time and contextualizes the contributions of each stage to the broader smart home landscape. 

Figure~\ref{fig:pub} reveals a consistent upward trend in publications, reflecting growing academic and industrial interest in smart home technologies. Notably, a significant surge in research activity is observed in the later years, coinciding with advancements in IoT and AI. Among the nine stages, contributions from Stages 6 (IoTs and Smart Home Ecosystems) and 8 (Sustainability and Energy Efficiency) dominate recent publications. This emphasizes the increasing focus on integrating interconnected ecosystems and promoting environmentally sustainable practices within smart home development. Additionally, the rise of Stage 9 (Brain-Computer Interfaces and Extended Reality) highlights the exploration of cutting-edge technologies, such as neural interfaces and immersive environments, which hold transformative potential for enhancing user experiences.

This analysis provides a comprehensive overview of the thematic and temporal progression of smart home research, highlighting areas where innovation is most active. The insights underscore the growing importance of sustainable, interconnected systems and the integration of advanced interfaces like extended reality and brain-computer technologies, guiding future efforts to tackle technological challenges while addressing user needs. In the next section, we delve into the evolution of smart homes, outlining the nine stages of their development to provide a structured understanding of their technological and conceptual progression.

\subsection{Evolution of Smart Homes}

\begin{figure}[!ht]
\centering
\includegraphics[width=1\linewidth]{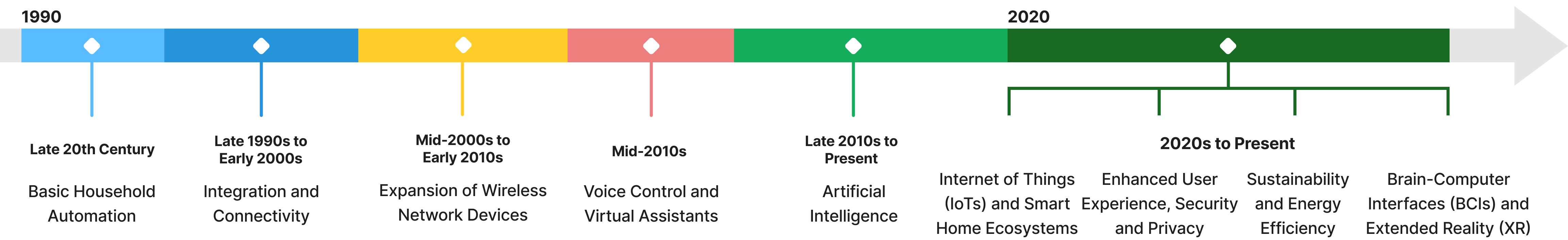}
\caption{\label{fig:review evolution}Evolution of smart homes.}
\Description{An image of smart homes evolution timeline.}
\end{figure}

The evolution of smart homes has been propelled by continuous advancements in science and technology, coupled with the increasing demand for greater household convenience \cite{ikezawa2020convenience}, energy efficiency \cite{salman2016energy}, and enhanced social connectivity \cite{dhelim2021iot}. Figure ~\ref{fig:review evolution} presents this progression in a timeline format, tracing the development from basic household automation in the late 20th century to the modern era dominated by the IoTs. To provide a clear understanding of the development of smart homes, we outline the key stages of their evolution in the following sections.

\subsubsection{Basic Household Automation (Late 20th Century) }
The concept of home automation \cite{aldrich2003smart}, introduced in the late 20th century, initially focused on basic tasks such as controlling lighting, thermostats, and electrical appliances through centralized systems \cite{bennett2017healthcare}. These early systems \cite{pirbhulal2016novel, jabbar2019design, sun2013multi} relied on physical wired connections and simple programming functionalities. The primary aim is to enhance convenience and comfort by automating repetitive household tasks. Although limited in scope, these early automation systems set the foundation for future innovations by demonstrating the potential of integrating technology into residential environments. Despite their high costs and the need for professional installation, they illustrated the transformative benefits of smart homes, paving the way for more scalable and user-friendly solutions in the years to come.

\subsubsection{Integration and Connectivity (Late 1990s to Early 2000s)}
With advancements in internet and telecommunications, smart home systems began incorporating remote connectivity and control capabilities. Broadband internet \cite{akerman2015skill} during this period played a crucial role, enabling devices to communicate more efficiently. The rise of mobile phones and early smartphones allowed homeowners to manage and monitor their homes remotely through dedicated web interfaces \cite{serrano2013mobile} or mobile applications \cite{serrano2013mobile}, significantly increasing the practicality of smart homes. This shift from isolated systems to networked devices marked a pivotal step in home automation, bringing the concept of remote monitoring into mainstream \cite{grubic2014servitization} use and making smart homes more adaptable, better acceptable and user-friendly.

\subsubsection{Expansion of Wireless Network Devices (Mid-2000s to Early 2010s)}
The mid-2000s marked a significant expansion in wireless smart home technology, thanks to the development of interoperability standards and the introduction of wireless protocols such as ZigBee \cite{han2014smart} and Z-Wave \cite{han2010smart}. These protocols provided standardized communication methods between devices, allowing for seamless integration of products from different manufacturers. Mesh networking \cite{benyamina2011wireless} further improved network reliability by enabling devices to act as repeaters, extending coverage throughout the home. These advancements not only reduced installation complexity and cost but also increased flexibility, allowing more homeowners to embrace smart home systems \cite{amaldi2008optimization}. The emphasis on interoperability was key in breaking the silos that previously existed between different manufacturers’ devices, making smart home adoption more accessible and effective.

\subsubsection{Voice Control and Virtual Assistants (Mid-2010s)}
The integration of voice control technologies, including Amazon Alexa, Google Assistant, and Galaxy Watch, introduced a new dimension of convenience and interactivity to smart homes. Voice assistants \cite{ammari2019music} transformed smart homes from manual, app-controlled environments to intuitive, hands-free experiences. Homeowners use simple voice commands to control lighting, temperature, entertainment systems, and even manage their daily schedules. The ability to receive personalized responses and automate tasks by speaking to a device added significant value, enhancing user engagement and making smart home technology more accessible, particularly for individuals with mobility issues or limited technical skills \cite{pal2020adoption}.

\subsubsection{Artificial Intelligence (Late 2010s to Present)}
The late 2010s saw the integration of AI into smart homes, significantly enhancing their capabilities. AI-driven devices learn from user behavior and preferences, enabling autonomous management of household tasks without explicit programming \cite{chang2021survey, zaidan2020review}. AI systems optimize energy usage \cite{zahraee2016application}, manage home security by detecting unusual activity, and adapt environmental conditions such as lighting and temperature based on occupant habits. This deep level of personalization makes smart homes more responsive to individual needs, resulting in an improved living experience \cite{aguilar2021systematic}. AI also allows smart homes to predict user needs and automate complex routines, further simplifying daily life and improving efficiency \cite{aguilar2021systematic}.

\subsubsection{IoTs and Smart Home Ecosystems (2020s to Present) }
The growth of IoT technologies has significantly expanded the capabilities of smart homes, allowing numerous appliances such as air conditioners, security cameras, door locks, and refrigerators to be interconnected within an integrated ecosystem \cite{choi2021smart}. Modern smart homes rely on a centralized controller that facilitates the seamless management of these interconnected devices. Cloud-based platforms \cite{fylaktopoulos2016overview} have also become integral, providing remote access, data storage, and enhanced capabilities such as real-time monitoring. This shift toward complete ecosystems represents the transition from isolated automation to an interconnected smart home experience, with devices working together to provide increased comfort, security, and energy efficiency \cite{li2019internet}.

\subsubsection{Enhanced User Experience, Security and Privacy (2020s to Present)}
As smart homes become more interconnected, robust security and privacy measures have become crucial \cite{yang2017survey}. Encryption \cite{zhou2013efficient}, secure transmission protocols \cite{saxena2014easysms}, and user authentication \cite{liu2017secure} are now foundational to protect sensitive user data. The implementation of privacy-by-design principles \cite{bu2020privacy} ensures that data collection is minimized, and users are given control over their personal information. Smart home manufacturers are also developing multi-layered security frameworks \cite{hong2016towards}, incorporating elements such as regular firmware updates \cite{zandberg2019secure}, anomaly detection \cite{chandola2009anomaly}, and intrusion prevention systems \cite{ierace2005intrusion} to ensure that smart home networks remain secure against potential cyber threats.

\subsubsection{Sustainability and Energy Efficiency (2020s to Present)}
Energy efficiency and sustainability have become central priorities in the development of smart homes. Smart energy monitoring systems, integrated renewable energy sources, and energy storage solutions enable homeowners to optimize energy consumption and minimize waste \cite{zhou2016smart}. Devices like smart thermostats \cite{adhikari2018algorithm}, energy meters \cite{sun2015comprehensive}, and AI-driven systems \cite{ali2024ai} help automate energy management, reducing consumption during peak hours and maximizing the use of renewable energy. This focus on sustainability not only reduces household costs but also contributes to broader environmental conservation efforts, making smart homes an essential component of green living \cite{sun2015comprehensive}.

\subsubsection{Brain-Computer Interfaces and Extended Reality (2020s to Present)}
The evolution of smart homes is accelerating with the integration of advanced technologies such as 6G \cite{dang2020should}, blockchain \cite{nofer2017blockchain}, edge computing \cite{shi2016edge}, and extended reality (XR) technologies \cite{ratcliffe2021extended}, including augmented reality (AR), virtual reality (VR), and mixed reality (MR), along with brain-computer interfaces (BCIs) \cite{kohli2022review, bhattacharya2021coalition}. BCIs offer the potential for users to interact directly with smart devices through neural commands, eliminating the need for traditional input methods. XR technologies provide immersive control interfaces, allowing users to visualize and interact with their smart home systems in novel ways \cite{kohli2022review}. These advancements are expected to significantly enhance not only the capabilities but also the user experience of smart homes, making them more interactive and deeply integrated into everyday human cognition and perception.

\subsection{Detection Technologies for Smart Homes}

\begin{figure}[!ht]
\centering
\includegraphics[width=0.8\linewidth]{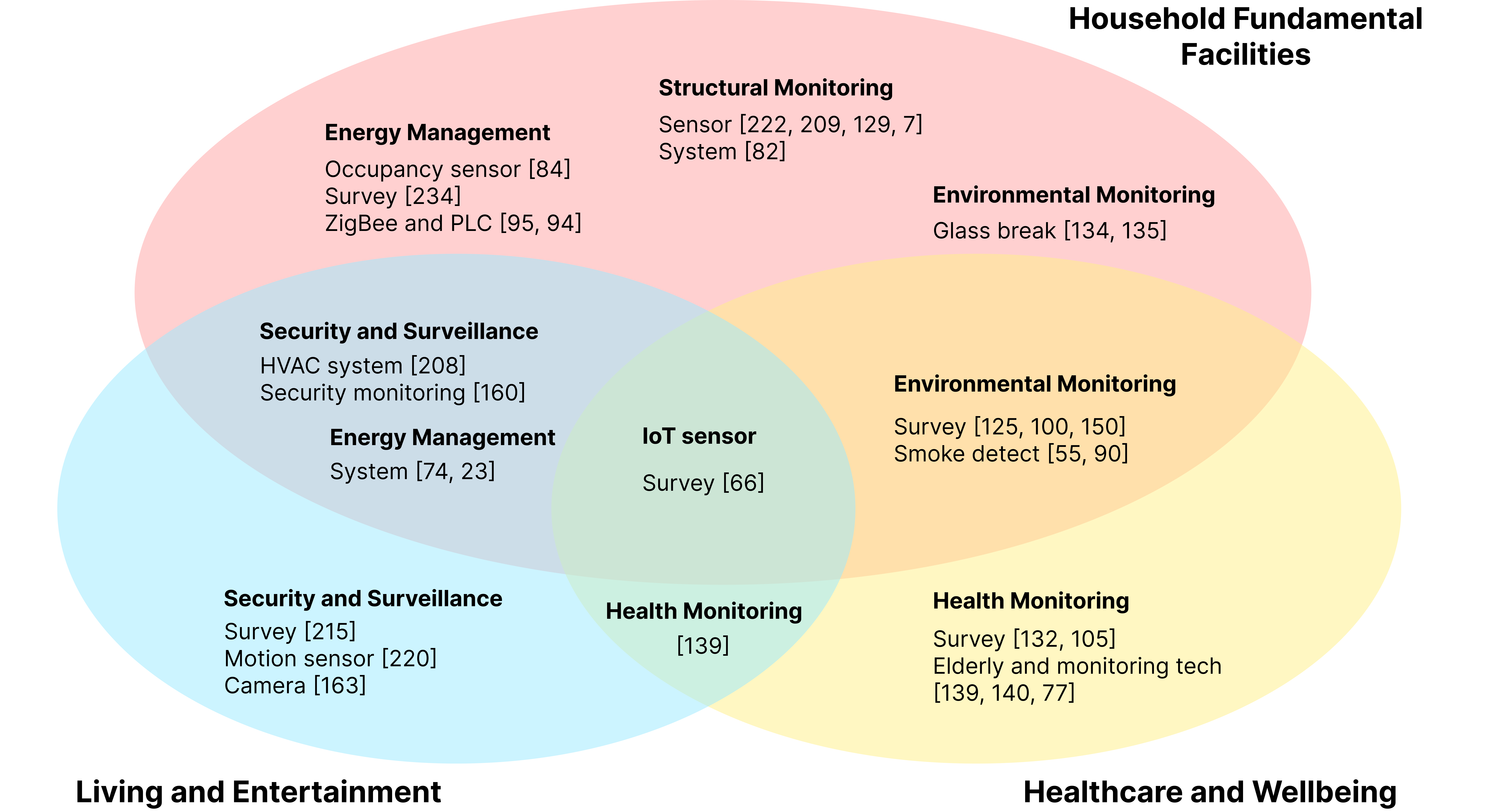}
\caption{\label{fig:key} 5 Key types of smarthome-based detection sensors.}
\Description{}
\end{figure}

Expanding on the trends identified in the previous analysis, smart home-based detection systems have become a cornerstone of innovation, particularly in stages 4 through 8 of smart home evolution, where technologies like IoT, AI, and interconnected ecosystems gained prominence. These systems and devices \cite{ding2011sensor, winkler2014security}, embedded within households, utilize advanced sensing, monitoring, IoT devices, AI, network, and automation technologies to detect and respond to various household events, activities, or changes. Designed to enhance privacy, security, convenience, energy efficiency, and overall functionality, these technologies create smarter, more efficient residential environments. Based on the literature reviews \cite{sun_systematic_2021, jiang2018smart, haroun2021progress}, Figure ~\ref{fig:key} highlights five key types of detection sensors widely used in smart homes, reflecting the technological advancements that drive these stages.

\subsubsection{Security and Surveillance} This category includes motion detectors, security cameras, and door and window sensors. AI-powered cameras now offer features like facial recognition, motion detection, and night vision, with advancements enabling them to distinguish residents from intruders and alert homeowners to potential security breaches.

Motion sensors \cite{wu2012classification} detect human activity in designated areas, enabling actions such as turning on lights, triggering alarms, or adjusting temperatures. Widely used in smart homes, these sensors include wearable, ambient, and multimedia types \cite{rashidi2012survey}, which are essential for activity recognition and ensuring safety. Innovations like electronic textiles offer lightweight, flexible, and cost-effective alternatives to traditional sensors, accommodating various body types while tracking real-time movements \cite{wang2020textile}. Advanced self-driven multi-functional motion sensors (MFMS) further enhance capabilities by measuring direction, speed, and acceleration using triboelectric nanogenerators and magnetic regulation modules \cite{narayana2018review, wu2018self}, making them versatile tools for smart home applications.

Smart home security cameras \cite{pierce2019smart} allow homeowners to monitor their property remotely, detecting motion, recording video, and sending alerts to devices when suspicious activity is observed. A near real-time monitoring and security system \cite{pandya2018smart} for smart homes using Wireless Sensor Networks (WSNs) \cite{akyildiz2010wireless} includes video cameras capable of identifying intruders, even if their faces are obscured by clothing, masks, or poor lighting conditions, with high accuracy \cite{pandya2018smart}.

Door and window sensors \cite{wu2015monitoring} detect the opening and closing of entry points, providing data on household entries and exits. These sensors are frequently used for security purposes and can also trigger automation routines, such as turning off the HVAC system \cite{trvcka2010overview} when a window is opened. A home automation system \cite{stolojescu2021iot} incorporating sensors and actuators linked via a flexible API, allowing users to control devices like door and window sensors remotely through smartphone apps. Another smart home monitoring system using ESP32 microcontrollers \cite{babiuch2020smart} integrates sensors and cameras with alarms and touchscreens, enhancing the security of doors and windows \cite{babiuch2020smart}.

\subsubsection{Environmental Monitoring} These technologies \cite{kumar2012environmental} detect changes in conditions such as smoke, carbon monoxide, radon, and other hazardous gases. They can alert residents to dangers even when they are away from home, and they can be integrated with other home systems to automate responses like turning on ventilation or contacting emergency services \cite{mois2017analysis}.

Smoke and carbon monoxide detectors \cite{chen2007fire} provide early warnings of potential hazards, sending alerts to homeowners' smartphones or triggering alarms within the house. These devices can also be connected with other systems to automatically shut off HVAC units, unlock doors, or notify emergency responders. \textit{Optimization of Smoke-Detector Installation} \cite{gu2023optimization} improves the placement of smoke detectors within homes to enhance detection efficiency. It emphasizes the importance of considering environmental factors like airflow, which can significantly influence the effectiveness of smoke detection and carbon monoxide absorption \cite{gu2023optimization}.

Glass break sensors detect \cite{mach2019accelerometer} sound or vibration patterns associated with breaking glass, alerting homeowners to potential intrusions or accidents like shattered windows. The accelerometer-based system \cite{mach2019accelerometer} distinguishes between actual intrusions and false alarms using an accelerometer to detect vibrations in glass panes. This technology \cite{mach2023development} increases security by accurately assessing glass break signals, aiming to meet stringent safety requirements.

\subsubsection{Energy Management} Energy management detection technologies \cite{zhou2016smart} in smart homes include smart thermostats and systems that monitor real-time electricity usage. These devices can identify patterns in energy consumption \cite{han2014smart} and automatically adjust heating, cooling, and lighting to improve efficiency and reduce costs \cite{anvari2014optimal}. 

Smart Home Energy Management Systems (SHEMS) leverage technologies like ZigBee and Power Line Communication (PLC) to monitor and control energy consumption and renewable energy generation \cite{zhou2016smart, han2014smart}. These systems integrate ZigBee-based measurement modules and PLC gateways to optimize energy usage, manage real-time data from appliances, and support renewable sources such as solar panels and wind turbines. Advanced features, including Disjoint Multi-Path Routing (DMPR) protocols, enhance network performance by efficiently routing sensor data, reducing energy costs through intelligent scheduling \cite{han2010smart}. Additionally, systems like solar-powered setups with Raspberry Pi enable remote appliance control, further improving energy efficiency and lowering operating costs in everyday use \cite{elkholy2022design}. Together, SHEMS promote sustainable energy practices while empowering homeowners with efficient energy management tools.

Occupancy sensors \cite{garg2000smart} play a key role in energy management by detecting whether rooms are occupied or vacant. These sensors can automate lighting, heating, and cooling systems to optimize energy usage based on the presence of occupants. They also provide security benefits by detecting whether a space is in use. Chaudhari et al. \cite{chaudhari2024fundamentals} thoroughly examine the underlying technologies used for occupancy detection, including the sensors and algorithms that support it. These technologies integrate into smart building systems, helping to improve energy efficiency, enhance security, and support emergency response strategies. Accurate occupancy data are effectively managing smart building environments, from lighting and HVAC control to security systems.

\subsubsection{Health Monitoring}
Health monitoring technologies \cite{huynh2018autonomous} in smart homes are particularly valuable for elderly care, offering devices capable of detecting falls, monitoring vital signs, and tracking movements within the home. Some advanced systems learn normal behavior patterns and can send alerts when unusual activities are detected, potentially signaling a health issue \cite{lynch2006summary}.

Unobtrusive health monitoring in private settings, such as homes, is facilitated by the use of advanced sensor technologies, big data, and AI, allowing for discreet health monitoring without invading privacy \cite{wang2021unobtrusive}. Smart homes designed for elderly healthcare \cite{majumder2017smart} integrate environmental and physiological sensors to support continuous remote health monitoring. This setup enables elderly individuals to live comfortably in their homes while being continuously monitored for potential medical emergencies. In addition, wearable sensors \cite{majumder2017wearable} integrated into smart home environments enable continuous tracking of vital signs, providing early detection of health issues and ensuring timely intervention.

Facchinetti et al. \cite{facchinetti2023can} further explore the role of smart home technologies in managing chronic diseases among older adults. Their study highlights the potential of these technologies to help control disease exacerbations and enhance patient safety by providing continuous health monitoring, thereby supporting disease management and timely interventions.

\subsubsection{Structural Monitoring}
Structural monitoring employs sensors to detect potential issues within a building's structure, such as shifts, cracks, floods, water leaks, or foundation problems. By identifying these integrity concerns at an early stage, this technology empowers homeowners to take proactive measures, preventing minor issues from escalating into significant damage and ensuring the long-term stability and safety of the home.

Structural sensors \cite{xu2004wireless} also monitor environmental factors like temperature, humidity, air quality, and noise levels, helping to maintain a comfortable indoor environment while identifying issues such as excessive humidity or poor air quality. Ullo, S. L. and Sinha, G. R. \cite{ullo2020advances} explore advancements in smart structural monitoring systems, highlighting the transformative role of IoT and modern sensor technologies. Their work emphasizes the monitoring of environmental factors such as air quality, water quality, and radiation pollution, demonstrating how IoT integration enhances the precision and effectiveness of structural management, enabling more efficient and proactive maintenance strategies.

An IoT-based solution for managing indoor environments in smart buildings \cite{floris2021iot} shows how IoT sensors measure parameters like temperature, humidity, air quality, and luminosity. The data collected from these sensors is integrated with machine learning to optimize building management, enhancing energy efficiency, occupant health, and overall comfort.

Smart structural sensors designed for water leak detection help homeowners prevent costly damage by identifying water where it shouldn’t be, such as around pipes, dishwashers, washing machines, or other potential leak sources. These sensors can be strategically placed in areas prone to water damage, such as basements, kitchens, and bathrooms, providing early warnings to prevent extensive damage. Water leak detectors monitor for signs of leaks or flooding, issuing alerts to homeowners that can mitigate potential water damage \cite{li2015review}. An integrated approach to leak detection using GIS and remote sensing has been applied in large-scale water distribution networks, offering methodologies such as infrared imaging and geographic information systems. While primarily focused on infrastructure, these techniques provide valuable insights that can be adapted for smart home water leak detection \cite{al2023integrated}.

These structural monitoring technologies are commonly integrated into centralized smart home systems, enabling seamless control through smartphone applications, voice assistants, or remote internet access. This integration facilitates real-time alerts, comprehensive data logging, and remote management capabilities, enhancing the safety, efficiency, and ease of managing modern homes. When combined with other smart home detection technologies, they contribute to an interconnected and intelligent ecosystem, offering homeowners improved control, convenience, and peace of mind. 

\begin{table*}[ht]
\scriptsize
\centering
\caption{Sensors commonly used in smart home detection technologies.}
\label{tab:top10 sensors}
\begin{tabular}{p{4cm}p{4.5cm}p{4.3cm}}
\hline
\textbf{Sensor Name}& \textbf{Functionality}& \textbf{Algorithm/Techniques}\\ 
\hline
IoT Environmental Sensors \cite{ullo2020advances} & Monitors air quality, temperature, etc. & Machine learning for optimization \\ 
Occupancy Prediction Sensors \cite{floris2021iot} & Detects the number of room occupants & Classification and regression models\\ 
Water Leak Detection Sensors \cite{tina2022water} & Detects water leaks in smart homes & IoT-based monitoring systems
\\ 
Smart Home Health Sensors \cite{wang2021unobtrusive} & Monitors health parameters unobtrusively & Semantic Web technologies\\ 
Air Quality Sensors \cite{ullo2020advances} & Monitors indoor air pollutants & IoT and AI for air quality analysis
\\ 
Smart Smoke Detectors \cite{ali2018cyber} & Detects smoke and fire hazards & IoT-enabled alert systems
\\ 
Carbon Monoxide Detectors \cite{pandya2018smart} & Detects harmful carbon monoxide levels & Electrochemical sensor technology
\\ 
Window and Door Sensors \cite{stolojescu2021iot} & Monitors and alerts on unauthorized entry & Machine learning for anomaly detection\\ 
Temperature and Humidity Sensors \cite{ullo2020advances} & Monitors environmental comfort levels & Statistical and predictive models\\ 
Noise Level Sensors \cite{marques2020real} & Monitors and manages indoor noise levels & Digital signal processing
\\ 
\hline
\end{tabular}
\end{table*}

Table ~\ref{tab:top10 sensors} provides an overview of the most commonly used sensors in smart home detection technologies, as identified in peer-reviewed literature. The discussion of detection technologies, including different types of sensors for smart homes, highlights their transformative potential in enhancing security, energy efficiency, and convenience, but their integration into daily life also brings forth critical ethical considerations. These technologies rely on extensive data collection and advanced algorithms to automate and optimize household functions, which raises concerns about privacy, fairness, and user autonomy. Beyond their technical performance, it is essential to evaluate how these technologies align with ethical principles, ensuring they empower users while safeguarding their rights and fostering trust. This transition underscores the need for a broader exploration of the ethical frameworks that guide the design and deployment of smart home systems, shifting the focus from technological capabilities to the human-centric and societal implications.

\subsection{Ethical Concerns in Smart Homes}

Exploring the ethical concerns surrounding smart homes, particularly in relation to health technologies for older adults, reveals several critical themes and challenges. Research emphasizes the importance of addressing privacy concerns, maintaining autonomy, and balancing technological solutions with human care \cite{pirzada_ethics_2022}.

A primary ethical issue involves privacy and data security. Smart home technologies often collect sensitive data, raising concerns about its usage and protection \cite{chung_ethical_2016}. Ensuring transparency and giving users control over their personal information is central to addressing these concerns. Unlike general AI systems, smart home technologies operate in deeply personal spaces, making their ethical implications more immediate and impactful. They should empower users rather than restrict their independence \cite{sanchez2017ethics}, which requires careful consideration of how these systems are deployed and the extent to which users are informed and able to control them \cite{chung_ethical_2016}. Another major ethical challenge is balancing technological assistance with human caregiving. While smart home technologies offer valuable support for health monitoring and daily living, they should complement, rather than replace, human interaction and care \cite{chung_ethical_2016}. Ideally, these technologies should foster human relationships instead of diminishing them, ensuring that they enhance quality of life without undermining social and emotional connections.

Studies \cite{chung_ethical_2016, sanchez2017ethics} provide in-depth reviews of the ethical implications of smart home technologies in elderly care, categorizing and analyzing the various concerns that must be addressed to ensure these solutions are effective and respect users' rights and dignity.

Current research on smart home ethics primarily focuses on four areas \cite{felber_mapping_2023}:

\begin{enumerate} 
\item \textbf{Privacy Concerns:} Privacy remains a significant ethical challenge as smart home technologies often involve the collection, storage, and sharing of extensive personal data. The primary concern lies in ensuring that users have control over their data and are fully informed about its usage. This is particularly crucial given the risks of data breaches and misuse, which can erode trust and compromise user security. Effective privacy safeguards and transparent policies are essential to protect user rights. 
\item \textbf{Autonomy and Independence:} Smart home technologies should enhance, rather than diminish, user autonomy. For older adults and individuals with disabilities, these systems must be designed to support independence while respecting decision-making abilities. Adaptive technologies should empower users without fostering over-dependence or undermining their sense of control over their environment.
\item \textbf{Human vs. Artificial Relationships:} A key ethical debate surrounds the role of technology in caregiving. While smart home systems can provide valuable support, they should complement rather than replace human interaction. Preserving and fostering human relationships alongside technological assistance is essential to ensure holistic care and to maintain the social and emotional well-being of users.
\item \textbf{Design and Inclusivity:} The design of smart home technologies must account for the diverse needs of users to avoid biases and ensure inclusivity. Accessibility should be a priority, accommodating people of varying abilities, languages, and cultural backgrounds. The challenge lies in creating solutions that address a broad spectrum of needs without perpetuating stereotypes or excluding certain user groups.
\end{enumerate}

Effectively addressing these ethical challenges demands a collaborative, interdisciplinary approach. Researchers, policymakers, industry leaders, and ethicists must work together to establish comprehensive guidelines and regulations that safeguard individual rights, promote inclusivity, and foster responsible innovation in smart home environments. This synergy is essential to ensuring that smart home technologies are both ethically sound and socially beneficial.

\section{AI Ethics From User Requirements' Perspective}

User requirements in smart homes are crucial for shaping personalized, user-friendly experiences, while also safeguarding privacy and security, promoting inclusivity, and fostering trust. Meeting these requirements encourages broader adoption of smart home technologies and supports their ethical and sustainable development. This section delves into AI ethics in smart homes from the perspective of user requirements, examining how technology can align with users' needs and expectations to create responsible, human-centered smart home ecosystems.

\subsection{Overview of URN}
The User Requirements Notation (URN) \cite{amyot2011user}, standardized by the International Telecommunication Union in 2008, is a leading modeling language for capturing and analyzing user requirements through goals and scenarios. URN is widely used in software engineering and system development processes, providing a structured approach to eliciting, documenting, and analyzing user requirements. This ensures effective communication between stakeholders and the development team, facilitating the design of systems that align with user needs.

URN integrates two key components: the Goal-oriented Requirement Language (GRL) \cite{amyot2010evaluating}, which focuses on modeling goals and objectives, and the Use Case Map (UCM) scenario notation \cite{amyot2000extension}, which models system behavior in various scenarios. Historically, these languages were supported by separate tools, but they are now integrated into a unified framework, enabling more comprehensive requirement analysis \cite{roy2006towards}. The current improvement of the URN consists of three main components: GRL, UCM, and Responsibility-Driven Design (RDD). Each of these components addresses specific aspects of requirements modeling, including goals, scenarios, and responsibilities.

GRL is used for modeling and reasoning about goals and non-functional requirements. It allows stakeholders to explore their concerns based on objectives, consider different alternatives to achieve those goals, and understand the impact of various decisions. GRL helps clarify what the system should accomplish and the rationale behind design choices, providing insight into the system’s goals and priorities \cite{roy2006towards}.

UCM focuses on functional requirements and scenarios. Mapping causal scenarios helps visualize workflows or processes within the system, linking responsibilities between different components. It provides a dynamic view of system functionality by showing how different system components interact and the paths that the system might take under varying conditions \cite{roy2006towards}.

RDD emphasizes assigning specific responsibilities to components or roles within the system. It helps clarify which entities are accountable for meeting particular requirements, ensuring transparency and responsibility throughout system design and implementation \cite{wirfs1990designing}.

Together, these components create a structured and systematic approach to requirements engineering, fostering collaboration, traceability, and validation throughout the development lifecycle. URN enables stakeholders to define, prioritize, and manage requirements effectively, leading to the successful delivery of software systems that meet user needs and expectations.

\subsection{Highlights of URN in Ethical Design}

URN plays a pivotal role in promoting ethical design by providing a structured framework that helps capture and analyze ethical considerations throughout the software development lifecycle. Below are key ways in which URN facilitates ethical design.

\subsubsection{Goal-Oriented Modeling} URN’s goal-oriented approach allows stakeholders to articulate and prioritize ethical goals and values within the software system. By explicitly modeling these ethical goals, developers can ensure that ethical considerations are incorporated from the outset of the design process \cite{amyot2010evaluating}. For instance, Bui-Thanh \cite{bui2007goal} presents a goal-oriented optimization framework that, although applied to dynamic models, can be extended to ensure software systems meet specific ethical objectives through carefully structured goals and constraints.

\subsubsection{Stakeholder Involvement} URN promotes the active involvement of diverse stakeholders, including end users, domain experts, and regulatory bodies, in the requirements elicitation process. This inclusive approach is essential for ethical design, as it allows for diverse ethical concerns and viewpoints to be considered. URN’s collaborative modeling capabilities foster transparency and inclusivity, ensuring ethical perspectives are embedded throughout the design process. Flechais and Sas \cite{flechais2009stakeholder} underscore the importance of stakeholder involvement in creating systems that balance usability and security, highlighting the value of diverse input in ethical system design.

\subsubsection{Traceability and Accountability} URN supports traceability mechanisms that link high-level ethical goals to detailed requirements and design decisions. This ensures accountability by allowing developers to demonstrate how ethical considerations are addressed throughout the development process. Additionally, URN enables impact analysis, allowing developers to evaluate the ethical implications of changes to the system. Ramesh \cite{ramesh2001toward} explores traceability in software development, showing how URN’s traceability features ensure that ethical goals remain visible and actionable throughout the project lifecycle.

\subsubsection{Scenario-Based Analysis} UCM provides visual representations of system behaviors, interactions, and scenarios, making it easier to identify and analyze ethical dilemmas and decision points. By modeling different ethical scenarios and exploring alternative actions, developers can make informed decisions aligned with ethical principles. Kazman et al. \cite{kazman1996scenario} demonstrate how scenario-based analysis simplifies the evaluation of systems, making it an effective tool for analyzing ethical consequences in software design.

\subsubsection{Responsibility Allocation} RDD approach helps developers allocate ethical responsibilities to specific components or actors within the system. This method promotes ethical accountability by ensuring that designated entities are responsible for key ethical concerns, such as privacy protection or mitigating biases in AI systems. Amyot \cite{amyot2010evaluating} notes that RDD aligns well with modern software engineering practices, emphasizing the need for systems to be adaptable and ethically sound, which is essential in maintaining accountability and addressing ethical risks.

URN provides a comprehensive, structured approach to integrating ethical considerations into software design. By leveraging its goal-oriented modeling, stakeholder involvement, traceability, scenario-based analysis, and responsibility allocation, developers can design systems that prioritize ethical values, respect user autonomy, and contribute to positive social outcomes.

\subsection{Implementations of URN in Smart Homes}

There are limited studies directly exploring the application of URN in smart homes, but the importance of capturing and analyzing user requirements remains evident in existing research. This paper searched for keywords "user requirements" and "smart homes" and identified several informative resources highlighting how URN can play a critical role in ensuring that smart home technologies are developed in alignment with user needs and ethical standards \cite{amyot2011user}.

Smart home systems typically integrate internet connectivity and sensor technologies to enhance automation, cost efficiency, and user convenience. Georgia et al. \cite{georgia2021evaluation} utilized the PairWise Comparison framework to prioritize user requirements in smart home services, with a focus on critical domains such as e-health and entertainment. Similarly, Brich et al. \cite{brich2017exploring} examined the challenges of designing intuitive smart home interfaces, particularly those that address both simple and complex user tasks. Through a contextual inquiry with 18 participants, their study found that while rule-based notations are effective for managing basic tasks, they lack the flexibility needed for more complex scenarios \cite{brich2017exploring}. These findings provide valuable insights for advancing future interface designs, ensuring they accommodate diverse user needs and improve overall usability.

In the context of smart homes, URN proves especially valuable for modeling complex scenarios involving interactions among multiple smart devices and various users operating under diverse conditions. By leveraging the GRL and UCM for scenario modeling, designers can effectively visualize and manage the intricate dynamics of the smart home ecosystem. This approach ensures that the system aligns with user preferences and fulfills specific requirements. Table ~\ref{tab:URN in smart homes} provides a summarized overview of the implementation of user requirements in smart home environments.

\begin{table*}[ht]
\scriptsize
\centering
\caption{URN in smart homes.}
\label{tab:URN in smart homes}
\begin{tabular}{p{2cm} p{5.2cm} p{5.8cm}} 
\hline
\textbf{Category} & \textbf{Description} & \textbf{Examples} \\ 
\hline
\textbf{Complex Scenario Modeling} & Uses various devices in scenarios for seamless operation. & Devices like smart bulbs and thermostats interact in settings like energy management. \\ 
\textbf{GRL}& Defines and prioritizes smart home goals into technical needs. & Adjusting temperature, optimizing energy, ensuring security. \\ 
\textbf{UCM}& Visual maps to show system behavior and identify issues. & Scenarios include coming home or when the house is empty. \\ 
\textbf{User-specific Requirements} & Addresses needs of all users, focusing on accessibility, customization, and privacy. & Features like accessibility for the  disabled, customization for household preferences, and privacy protections.\\ 
\hline
\end{tabular}

\end{table*}

\subsubsection{Complex Scenario Modeling}
Smart home technology integrates various devices such as smart bulbs, thermostats, security cameras, and voice assistants, all of which must work together seamlessly in different scenarios. URN allows designers to create detailed models that show how these devices interact with users and each other across multiple use cases. This helps ensure that the system performs reliably under real-world conditions, addressing the complexity of interactions between devices and user behaviors.

\subsubsection{Goal-oriented Requirement Language}
This enables designers to define and prioritize goals and requirements for smart home systems. For instance, users may desire automatic temperature adjustments, energy optimization, and robust security measures. GRL helps break down these high-level goals into actionable technical requirements, ensuring that user preferences are clearly articulated and systematically addressed during development.

\subsubsection{Use Case Map}
This component of URN provides a visual representation of system behaviors and interactions, allowing designers to uncover potential conflicts, dependencies, and redundancies. By modeling various user scenarios—such as "arriving home in the evening" or "leaving the house unattended"—UCM ensures that the smart home system operates according to user needs across different contexts, enhancing its adaptability and usability.

\subsubsection{User-specific Requirements}
Smart home systems must consider the diverse needs of all users, including the elderly and disabled. URN facilitates the inclusion of special requirements during the design phase, ensuring systems are accessible and customizable to fit unique household needs. With URN, designers can offer flexible customization options, allowing users to personalize system settings based on individual preferences, improving user experience and satisfaction.

Privacy is a critical concern in smart home environments. URN enables designers to model and analyze privacy requirements, ensuring that sensitive user data is protected throughout the system. For instance, URN can be used to model the camera's field of view or data storage strategies, ensuring that private spaces are not unnecessarily monitored and that data is handled securely.

\subsubsection{Mitigating Privacy and Bias Risks}
URN’s modeling capabilities allow designers to identify and mitigate privacy risks posed by detection technologies during the design phase \cite{amyot2011user}. For example, URN can be used to assess the field of view of cameras or the strategies for handling recorded data, ensuring private areas are not inadvertently monitored. By integrating encryption and access control measures into the design, URN helps safeguard user data against unauthorized access.

When designing detection technologies, URN also helps identify potential biases. For example, sound recognition devices may struggle to accurately interpret certain accents or languages. By modeling these issues, URN allows designers to anticipate and resolve bias-related challenges. URN further provides a structured approach to evaluating the performance of detection technologies across diverse user groups, ensuring fairness and accessibility for all users.

\subsubsection{Responsibility Allocation}
Using the RDD method, designers can clearly assign responsibilities for maintaining privacy and security within the system. This clarity ensures that if issues arise, the responsible parties can be quickly identified and held accountable, allowing for prompt corrective actions. URN fosters transparency and accountability, defining clear roles for all stakeholders—users, developers, and regulatory bodies—in the design and implementation of the system.

URN provides a comprehensive and structured framework for developing smart home technologies that not only satisfy technical specifications but also prioritize critical ethical considerations such as privacy, security, fairness, and inclusivity. By harnessing the capabilities of URN, developers can create smarter, safer, and ethically responsible systems that are thoughtfully aligned with user needs and societal values, ensuring a balance between innovation and ethical accountability.

\subsection{URN for Ethical Consideration in Design of Detection Technologies}

Incorporating URN into the ethical evaluation of smart home technologies enables the systematic mapping and analysis of the ethical implications of various design choices. URN facilitates the early identification of potential ethical issues, such as data misuse, lack of transparency, or infringements on user autonomy, during the initial stages of the design process. By embedding ethical considerations directly into the requirements and design phases, URN ensures that these concerns are integral to the technology's core functionality rather than treated as secondary considerations. This proactive approach fosters the development of smart home technologies that are not only technically proficient but also ethically sound and user-centric, aligning with the growing demand for responsible and transparent innovation \cite{chung_ethical_2016}.

The application of URN has extended to the design of specific detection technologies in smart homes, including motion detectors, cameras, and sound recognition devices. Using URN, designers can model the interactions between these technologies and other smart home components, ensuring that they comply with user privacy and security requirements. URN provides a framework for identifying scenarios where detection technologies might infringe on privacy or introduce biases, enabling developers to address these issues proactively by redesigning the technology or implementing additional safeguards.

For instance, motion detectors are frequently used in smart homes for security and automation purposes \cite{van2017design}. URN allows designers to model various motion detection scenarios—such as differentiating between family members, pets, or intruders—minimizing the risk of false alarms. Additionally, URN can help analyze the sensitivity and coverage of motion detectors, optimizing their placement to avoid monitoring private spaces and ensuring privacy protection \cite{borst1989principles}.

Cameras play a significant role in smart homes for purposes ranging from security surveillance to video communication. With URN, designers can create models that map how cameras capture, store, and transmit video data, ensuring compliance with privacy regulations. URN can also help identify potential risks, such as unauthorized access or data leaks, and propose solutions such as encryption and strict access controls to enhance security \cite{bae2011product}.

Sound recognition devices, used for voice control and environmental monitoring, can also be modeled with URN to simulate various use cases, including voice command recognition and emergency detection. URN enables designers to evaluate factors like misrecognition rates and response times, allowing for improvements that ensure these devices perform reliably without compromising user privacy \cite{venkatraman2021smart}.

From an ethical perspective, URN addresses three aspects in the design of detection technologies:

\begin{itemize}
    \item \textbf{Transparency and Accountability:} URN’s modeling capabilities allow developers to trace and document how ethical issues are managed throughout the design and development stages \cite{felzmann2019transparency}. This includes modeling the data collection and usage processes to ensure compliance with privacy regulations and maintaining transparency in how user data is handled.
    \item \textbf{Stakeholder Engagement:} URN facilitates the inclusion of various stakeholders—such as end-users, domain experts, and regulatory bodies—in the design process \cite{hasselqvist2018designing}. Collaborative modeling sessions provide opportunities for stakeholders to voice their ethical concerns, ensuring that their perspectives are considered, and fostering transparency and inclusivity.
    \item \textbf{Responsibility Allocation:} The RDD of the URN approach helps allocate specific ethical responsibilities to various actors or components within the system. By assigning clear responsibilities, URN promotes ethical accountability and reduces the risk of unethical practices, such as privacy breaches or biased decision-making \cite{chu1979adaptive}.
\end{itemize}
    Through these features, URN offers a structured approach to ensuring that the design of detection technologies in smart homes is not only technically robust but also ethically aligned with user expectations and regulatory standards.

\subsection{Design Process and Evaluation Guidelines following URN}

Next, we address the key ethical considerations in the design and evaluation of detection technologies, focusing on critical themes such as privacy and data protection, bias and fairness, transparency and explainability, accountability and responsibility, and human autonomy and decision-making.

\subsubsection{Privacy and Data Protection}
The integration of detection technologies into smart homes raises significant concerns about privacy and data protection. Devices such as smart cameras, voice assistants, and motion detectors collect extensive amounts of sensitive data, including visual, auditory, and behavioral patterns of residents \cite{muslukhov2012understanding}. To mitigate these risks, robust data protection mechanisms must be employed, including encryption for both data transmission and storage. Furthermore, clear policies are needed to regulate data access and sharing, ensuring that data is used exclusively for its intended purposes. Users should maintain control over their personal data through transparent consent mechanisms and accessible privacy settings.

The Privacy by Design \cite{perera2016privacy} framework is instrumental in embedding privacy considerations into the technology's development from the outset, rather than as an afterthought. This approach advocates for data minimization, collecting only the data necessary for specific purposes and ensuring encryption throughout its lifecycle. Additionally, default privacy settings should be set at the highest level, allowing users to modify them as needed. By integrating these principles, Privacy by Design not only safeguards user data but also fosters trust by demonstrating a commitment to privacy across the product lifecycle.

\subsubsection{Bias and Fairness}
Detection technologies, including facial recognition and behavior-tracking algorithms, are prone to perpetuating biases if not carefully managed. These biases can arise from imbalanced training datasets or flawed algorithmic designs, leading to discriminatory outcomes based on race, gender, or socioeconomic status  \cite{mehrabi2021survey}. To address these issues, developers must ensure the use of diverse datasets for training and regularly test algorithms to detect and mitigate biases. Mechanisms should be established to promptly correct any unfair outcomes that may occur.

Ensuring fairness in algorithmic decision-making \cite{starke2022fairness} is vital to prevent the amplification of existing social biases. In smart homes, this includes algorithms governing everything from energy management to security systems. Developers should adopt techniques to identify and mitigate biases during both the data preparation and algorithm development stages. This can be achieved through regular fairness audits, the use of unbiased datasets, and the implementation of models capable of dynamically adjusting to prevent discriminatory outcomes.

\subsubsection{Transparency and Explainability}
Transparency and explainability \cite{balasubramaniam2022transparency} are essential for fostering user trust, particularly in autonomous systems. Users need to be informed about what data is being collected, how it is processed, and how decisions are made by smart home technologies \cite{balasubramaniam2022transparency}. For instance, if an anomaly detection system triggers an alert, the user should be provided with a clear explanation of the alert and the reasoning behind the system's decision. This transparency enables users to make informed choices about privacy and security settings.

In smart home technologies, transparency involves clearly communicating how, why, and what data is collected, as well as how it is processed. Explainability further ensures that users can understand the rationale behind the decisions made by AI systems. This might involve developing user interfaces that provide concise explanations of system actions. For example, if a smart thermostat automatically adjusts the temperature, the system should offer a clear explanation via a user interface or app. These measures are essential for building trust and enabling users to manage their privacy and security settings effectively.

\subsubsection{Accountability and Responsibility}
Accountability in smart home detection technologies requires clearly defining who is responsible for the actions and outcomes of these systems \cite{smith2021clinical}. Developers must be held accountable for the performance and ethical compliance of their products. This includes being responsible for any data breaches, misuse of technology, or unintended consequences that arise from their systems. Establishing regulatory frameworks is crucial to ensure that companies adhere to ethical standards.

Robust accountability mechanisms are essential to ensure that designers of smart home technologies are held responsible for the ethical implications of their products. This requires the establishment of clear standards and regulations to address issues such as privacy breaches, biased decision-making, and misuse of data. Furthermore, user feedback channels should be implemented, enabling individuals to report concerns directly to regulatory bodies or relevant authorities. Such mechanisms not only safeguard user rights but also foster transparency and drive the continuous improvement of smart home technologies, promoting ethical innovation and trust.

\subsubsection{Human Autonomy and Decision-Making}
Preserving human autonomy is crucial as smart home technologies become more automated. These technologies should enhance, rather than diminish, users' decision-making capabilities \cite{barnes2017humans}. Systems should be designed to assist users, allowing them to intervene and retain final control over decisions. For instance, while automated systems may suggest or pre-configure settings based on user behavior, the ultimate decision should always rest with the user, with simple options available to override the system's decisions.

Maintaining human autonomy necessitates designing smart home technologies that empower users with full control over their environment. This involves ensuring that significant decisions are never made without the user’s explicit consent and that users can effortlessly override automated actions when needed. Achieving this requires transparent disclosure of device functionalities and data processing practices to secure informed consent. For instance, while systems can suggest or pre-configure settings based on user behavior, the ultimate decision must rest with the user, supported by intuitive and readily accessible options to modify or override these automated configurations.

These ethical guidelines aim to ensure that smart home technologies better protect user privacy, promote fairness, and build trust. By addressing these considerations, smart home systems can integrate more ethically into daily life, fostering responsible and user-centric innovation.

\section{AI Ethics From Technology's Perspective}
The discussion on AI ethics from user requirements perspectives highlights the importance of designing smart home technologies that prioritize inclusivity, transparency, and user autonomy. Building on this foundation, it is equally critical to explore how these ethical principles translate into the technological frameworks that underpin smart home systems. As IoT and AI technologies continue to redefine the capabilities of smart homes, the focus shifts to examining the algorithms and models driving these innovations. These technological advancements, while enabling unprecedented levels of connectivity, efficiency, and personalization, also introduce complex ethical challenges that demand rigorous analysis and proactive solutions.

\begin{table*}[htbp!]
\centering
\tiny
\caption{Overview of IoT technologies in peer-reviewed papers.}
\label{tab:iot_AI_ethics}
\begin{tabular}{p{0.7cm}p{0.2cm}p{0.3cm}p{0.3cm}p{2.2cm}p{2.2cm}p{2.2cm}p{1.6cm}p{0.3cm}p{0.3cm}}
\hline
\textbf{Reference} & \textbf{IoT} & \textbf{Smart home} & \textbf{AI ethics} & \textbf{Research field} & \textbf{Algorithm}  &\textbf{Model}& \textbf{Platform}  & \textbf{Year} &\textbf{Cite} \\ 
\hline
\cite{cook2012casas} & \ding{51} & \ding{51} & \ding{55} & Pervasive Computing, Machine Learning& Support Vector Machines (SVM) &Activity recognition and discovery model
& CASAS SHiB Kit  & 2013 
&906 
\\ 
\cite{hussain2020machine} & \ding{51} & \ding{51} & \ding{55} & IoT, Security, Deep Learning& SVM, Random Forest, Naive Bayes, KNN, DNN  &IoT Security Models
& IoT Networks  & 2020 
&825 
\\ 
\cite{han2010design} & \ding{51} & \ding{51} & \ding{55} & Energy Management, ZigBee, Sensor Network& DMPR (Disjoint Multi Path Routing Protocol)  &Active sensor network model for energy management.
& ZigBee-based sensor networks  & 2010 
&781 
\\ 
\cite{zhang2020empowering} & \ding{51} & \ding{51} & \ding{51} & IoT, Deep Learning, Edge Computing, Fog Computing& Deep Neural Networks (DNNs), Reinforcement Learning (RL)  &AIoT System Models
& IoT Devices, Edge Servers, Cloud Platforms  & 2020 
&598 
\\ 
\cite{shafiq2020corrauc} & \ding{51} & \ding{55} & \ding{55} & IoT, Malicious Traffic Detection, Machine Learning & CorrAUC, C4.5 Decision Tree, Random Forest, SVM, Naive Bayes  &CorrAUC feature selection model
& IoT networks, Bot-IoT dataset  & 2021 
&504 
\\ 
\cite{merenda2020edge} & \ding{51} & \ding{55} & \ding{55} & Edge Computing, Machine Learning& Deep Learning, CNN, SVM, KNN  &
& IoT devices, Embedded devices, Edge servers  & 2020 
&439 
\\ 
\cite{mohammadi2018enabling} & \ding{51} & \ding{51} & \ding{51} & Big Data, Machine Learning& Deep Reinforcement Learning (DRL)  &Cognitive Smart City Framework
& IoT Devices, Smart City Infrastructure  & 2018 
&414 
\\ 
\cite{yu2019deep} & \ding{51} & \ding{51} & \ding{55} & Energy Management& Deep Reinforcement Learning (DRL)  &Markov Decision Process (MDP)-based optimization for energy management.
& IoT-enabled smart home systems  & 2019 
&369 
\\ 
\cite{sarker2023internet} & \ding{51} & \ding{55} & \ding{55} & IoT Security, Machine Learning
& SVM, RF, K-means  && IoT devices, Cloud systems  & 2023 
&309 
\\ 
\cite{ali2020state} & \ding{51} & \ding{51} & \ding{55} & AI, Smart Grid & RL, ANN, GA  &Security Intelligence model, Deep Learning architectures
& Smart Grid  & 2020 
&278 
\\ 
\cite{celik2017decentralized} & \ding{51} & \ding{51} & \ding{55} & Energy Management Systems & Genetic Algorithm  &Multi-agent system-based optimization for coordinated energy management.
& Renewable energy  & 2017 
&194 
\\ 
\cite{kishor2022artificial} & \ding{51} & \ding{51} & \ding{51} & AI, IoT, Healthcare & Decision Tree (DT), Support Vector Machine (SVM),Naïve Bayes (NB)&AI and IoT-based healthcare prediction model
& IoT-enabled healthcare system  & 2022 
&189 
\\ 
\cite{gonzalez2019review} & \ding{51} & \ding{51} & \ding{51} & AI, IoT & Random Forest  &AI-Integrated IoT Models
& IoT Devices  & 2019 
&182 
\\ 
\cite{djenouri2019machine} & \ding{51} & \ding{51} & \ding{55} & Smart Buildings & PCA, K-Means  &Smart Building ML Models
& IoT, Smart Buildings  & 2019 
&172 
\\ 
\cite{alberdi_smart_2018} & \ding{51} & \ding{51} & \ding{51} & Alzheimer's Prediction, Behavior Monitoring& Vector Regression, AdaBoost, Random Forest, Multilayer Perceptron&ML models for predicting Alzheimer's symptoms based on in-home behavior& IoT sensor networks  & 2018 
&170 
\\ 
\cite{el2021secure} & \ding{51} & \ding{51} & \ding{51} & Healthcare Monitoring& Kerberos authentication  &Fuzzy Logic-based Healthcare System Model
& Wearable sensors  & 2021 
&154 
\\ 
\cite{baccour2022pervasive} & \ding{51} & \ding{51} & \ding{51} & Distributed AI & FedAvg  &Pervasive AI Model for Distributed IoT Systems
& IoT Devices  & 2022 
&117 
\\ 
\cite{aversano2021systematic} & \ding{51} & \ding{51} & \ding{55} & IoT Security & CNN, RNN, DQN, GAN, DBN, Autoencoders, SVM
&Deep Learning-based IoT Security Model
& IoT Devices & 2021 &108 \\ 
\hline
\end{tabular}
\end{table*}

Table ~\ref{tab:iot_AI_ethics} provides a detailed overview of influential IoT technologies, selected based on their relevance, contributions, and citation counts to key research areas. These highly cited studies represent significant advancements across domains such as energy management \cite{han2010design, celik2017decentralized}, healthcare \cite{kishor2022artificial, el2021secure}, and security \cite{hussain2020machine, aversano2021systematic}, showcasing their versatility and impact in addressing diverse user needs. The selected works highlight not only the technical capabilities of IoT solutions but also their potential to integrate ethical principles into their frameworks.

Notably, some studies explicitly address AI ethics, tackling concerns like fairness, privacy, and accountability. For instance, research incorporating ethical considerations includes models focused on healthcare monitoring and distributed AI \cite{kishor2022artificial, baccour2022pervasive}. This growing emphasis on ethical alignment reflects an increasing awareness of the need to balance technological innovation with societal values, ensuring smart home systems are both functional and ethically sound.

Advanced techniques like reinforcement learning \cite{kaelbling1996reinforcement}, deep neural networks \cite{sze2017efficient}, and federated learning \cite{li2020federated} emerge as prominent tools for optimizing system performance while safeguarding sensitive user data. For instance, applications in healthcare and energy optimization aim to balance technical efficiency with ethical considerations, ensuring equitable outcomes and robust data protection. However, the analysis also reveals gaps, with many studies overlooking explicit ethical frameworks, underscoring the necessity for broader adoption of principles that prioritize inclusivity, transparency, and user autonomy. This comprehensive overview underscores the imperative of harmonizing ethical considerations with technological advancements to foster smarter, safer, and more equitable smart home ecosystems.

The adoption of smart home-based detection technologies introduces complex ethical considerations that demand thorough investigation. While these technologies promise significant improvements in home security, health monitoring, and convenience, they also pose challenges related to privacy, informed consent, and user autonomy \cite{chung_ethical_2016}. This section delves into these ethical challenges from a technological perspective, examining the latest developments in algorithms and models, cybersecurity and networks, and hardware and software, offering a holistic view of the intersection between innovation and ethical responsibility.

\subsection{Latest algorithms and models}
AI ethics are increasingly integrated into designing and implementing the latest algorithms and models. The ways AI ethics manifest in smart home-based detection technologies include algorithmic fairness \cite{kleinberg2018algorithmic}, diverse datasets \cite{majumdar2021security}, explainable AI (XAI) \cite{arrieta2020explainable}, federated learning \cite{zhang2021survey}, adversarial robustness \cite{carlini2019evaluating} and secure AI \cite{yu2021secure}.

\subsubsection{Algorithmic Fairness}
Fairness-aware algorithms are designed to detect and mitigate biases in both data and model predictions, ensuring equitable outcomes across different demographic groups. Techniques such as re-sampling, re-weighting, and applying fairness constraints are commonly used to promote fairness in algorithmic decision-making. In the rapidly evolving landscape of smart home technologies, human-in-the-loop IoT systems are increasingly able to autonomously adapt to human and environmental conditions. However, the inherent variability in human behavior presents a significant challenge to maintaining algorithmic fairness. To address this, FaiR-IoT \cite{elmalaki2021fair}, a reinforcement learning-based framework, enhances personalization, performance, and fairness. This framework has demonstrated success in applications like smart homes, leading to substantial improvements in both fairness and overall system performance \cite{pirzada_ethics_2022, zhu_ethical_2022}.

The concept of fairness in Demand Side Management (DSM) systems \cite{baharlouei2014efficiency} introduces optimal billing mechanisms that equitably distribute energy costs based on individual user contributions to minimize overall system costs in smart homes. This approach ensures fairness by adhering to resource allocation principles, while also protecting user privacy through the implementation of the secure sum algorithm \cite{birchley_smart_2017}. This mechanism not only enhances cost efficiency but also upholds fairness and privacy within energy management systems.

\subsubsection{Diverse Datasets}
To minimize inherent biases, training data is curated to ensure it includes diverse and representative samples. Regular dataset audits are conducted to identify and correct imbalances, ensuring fairness in model predictions. In addressing the security challenges of IoT devices, Majumdar et al. \cite{majumdar2021security} presents a methodology that extracts actionable security rules from established standards and best practices, applying them to low-level device data to conduct thorough audits. This methodology is evaluated for its efficiency and scalability within smart home environments, highlighting the critical role that diverse datasets and rigorous auditing play in ensuring robust IoT security.

\subsubsection{Explainable AI}
New models and frameworks are being developed to enhance the interpretability of AI decisions, making them more understandable to humans. Techniques such as Local Interpretable Model-agnostic Explanations (LIME), SHapley Additive exPlanations (SHAP), and Anchors provide valuable insights into how models reach their conclusions \cite{zhou2021evaluating}. In the domain of remote caregiver monitoring, a study evaluated both machine learning experts and non-experts, finding that SHAP-generated explanations were 92\% accurate, with 83\% of users preferring detailed explanations over simple activity labels \cite{das2023explainable}. While certain explainable AI (XAI) methods can enhance user trust in activity recognition models, others may undermine it, highlighting the ethical importance of selecting the appropriate XAI approach \cite{zhou2021evaluating}. Furthermore, in sensor-based recognition of Activities of Daily Living (ADL) within smart home environments—crucial for healthcare and ambient assisted living—XAI plays a key role in fostering trust and transparency. The DeXAR framework \cite{arrotta2022dexar} converts sensor data into semantic images, allowing the application of XAI methods based on Convolutional Neural Networks (CNN) \cite{wu2017introduction}, which then generate natural language explanations. Among these methods, white-box XAI, using prototypes, proved to be the most effective in ensuring ethical transparency and trustworthiness in ADL recognition systems, reinforcing the importance of ethical design in smart home technologies \cite{arrotta2022dexar}.

\subsubsection{Federated Learning}
This approach enables models to be trained across multiple devices or servers that hold local data samples, without requiring the exchange of raw data, thus enhancing data privacy. FedHome \cite{wu2020fedhome}, an ethical cloud-edge federated learning framework that preserves user privacy by keeping data local while building a shared global model. It utilizes a generative convolutional autoencoder to deliver precise and personalized health monitoring, along with efficient cloud-edge communication to manage imbalanced and non-identically distributed data. Experiments using realistic human activity recognition datasets demonstrate that FedHome outperforms existing methods, underscoring the ethical importance of scalable, privacy-preserving in-home health monitoring systems \cite{wu2020fedhome}. Additionally, LoFTI \cite{yu2020learning} introduces a federated multi-task learning framework designed to ethically and automatically learn contextual access control policies from user behavior in smart homes. It addresses challenges such as data insufficiency, diversity, and privacy through a privacy-preserving federated approach. By capturing contextual access patterns using six identified feature types and employing a ML model with temporal structure and custom data augmentation, LoFTI significantly reduces false negatives (by 24.2\%) and false positives (by 49.5\%) compared to traditional methods. This highlights the critical role of ethical AI in developing secure, privacy-conscious smart home systems \cite{yu2020learning}.

\subsubsection{Adversarial Robustness}
Models are increasingly designed to be robust against adversarial attacks that could manipulate inputs to produce incorrect or harmful outputs, ensuring reliability and security in AI systems. ADDetector \cite{li2021federated}, a privacy-preserving Alzheimer's detection system that employs advanced security measures to safeguard sensitive data. ADDetector collects audio data and extracts topic-based linguistic features to improve diagnostic accuracy. It operates using a three-layer architecture—user, client, and cloud—to ensure data privacy. The system integrates federated learning (FL) and differential privacy (DP) to maintain data confidentiality and integrity, and it utilizes an asynchronous, privacy-preserving aggregation framework to secure model updates. Tested on 1,010 trials from 99 users, ADDetector achieved an 81.9\% accuracy rate with just a 0.7-second time overhead. This highlights the ethical importance of privacy-preserving AI, particularly in sensitive healthcare applications where user data security and model accuracy are paramount.

\subsubsection{Secure AI}
Implementing secure coding practices and conducting regular security assessments are essential for protecting AI systems from vulnerabilities and threats. The Anonymous Secure Framework (ASF) \cite{kumar2017anonymous} addresses ethical concerns related to security and privacy by utilizing lightweight operations that enhance security with minimal user intervention. ASF provides efficient mechanisms for authentication, key agreement, device anonymity, and unlinkability, using a one-time session key progression to reduce the risk of compromised keys. This framework offers lower computational complexity while improving security compared to existing schemes, underscoring the ethical imperative for robust, privacy-preserving mechanisms in smart homes \cite{kumar2017anonymous}.

To tackle security challenges in smart home systems, another framework combines secure AI and blockchain technology \cite{shah2023ai}. This framework utilizes the Isolation Forest (IF) \cite{staerman2019functional} algorithm to filter anomalies from a standard smart home dataset, followed by the application of classification algorithms such as k-nearest Neighbors (KNN) \cite{guo2003knn}, Support Vector Machines (SVM) \cite{joachims1998making}, Linear Discriminant Analysis (LDA) \cite{lu2003face}, and Quadratic Discriminant Analysis (QDA) \cite{scikit-learn_lda_qda} to distinguish attack data from non-attack data. To safeguard the integrity of the data, the Interplanetary File System (IPFS) securely stores non-attack data, with its hash recorded on the blockchain’s immutable ledger to prevent data manipulation. This framework demonstrated high performance, with KNN achieving 99.53\% accuracy during training and IF reaching 99.27\% accuracy in anomaly detection, alongside strong validation and scalability metrics. These results showcase its effectiveness in enhancing IoT security and ensuring the ethical implementation of AI in smart homes \cite{shah2023ai}.

Integrating AI ethics into the development of algorithms and models marks a significant advancement in creating smart home technologies that are fairer, more transparent, and secure. As we move forward, a continued emphasis on strengthening cybersecurity and network infrastructure will be crucial in safeguarding these innovations, ensuring that smart homes remain both safe and resilient against emerging cyber threats and vulnerabilities.

\subsection{Cyber and networks}
Advancements in AI ethics play a pivotal role in shaping smart home-based detection technologies, ensuring they are fair, transparent, secure, and privacy-preserving. As we continue integrating these ethical principles into algorithms and models, addressing the challenges posed by cybersecurity and network vulnerabilities in smart home systems becomes equally critical. The complexity of modern smart homes, characterized by interconnected devices and vast data streams, requires robust cybersecurity measures to protect against potential threats and system vulnerabilities.

Machine learning-based Intrusion Detection Systems (IDS) \cite{khraisat2019survey} offer effective methods for detecting cyberattacks in IoT networks. However, they also introduce a new vulnerability: Adversarial Machine Learning (AML) \cite{anthi2021hardening}. AML can manipulate data and network traffic, causing malicious packets to be misclassified as benign. This delay in attack detection poses risks such as personal data breaches, hardware damage, and financial loss. A rule-based method was proposed to generate AML attack samples and assess their impact on supervised classifiers for detecting Denial of Service (DoS) \cite{anthi2021hardening} attacks in smart home networks. By identifying key DoS packet features for perturbation, the study demonstrates how adversarial samples, combined with adversarial training, can strengthen IDS robustness. This underscores the importance of incorporating ethical AI practices to enhance the reliability and security of IDS systems amid evolving threats like AML.

Smart homes also face physical threats from advanced cyber-attacks and cyber-physical system attacks \cite{sapalo2019vpnfilter}. Defense mechanisms are developed to address a range of cyber threat attacks, particularly focusing on efficient task management and an in-depth understanding of VPNFilter \cite{sapalo2019vpnfilter} malware attacks, a significant threat to IoT devices.

IoT-based smart home solutions are highly susceptible to data leakage, motivating researchers to explore more secure alternatives. One significant threat is wireless signal eavesdropping, where attackers can intercept sensitive information even if communications are encrypted \cite{nassiri2022smart}. An example of such a threat is the "fingerprint and timing-based snooping (FATS)" attack \cite{nassiri2022smart}, a side-channel attack (SCA) \cite{nassiri2022smart} that passively infers in-home activities from a remote location by analyzing timing and fingerprint patterns. SCAs exploit valuable information from smart systems without accessing the actual data packet content. Nassiri et al. \cite{nassiri2022smart} provide a comprehensive review of SCAs in cyber-physical systems, focusing on solutions to mitigate FATS attacks.

These proactive approaches to addressing cybersecurity and network vulnerabilities, within the framework of AI ethics, are crucial for improving the security, reliability, and trustworthiness of smart home technologies.
 
\subsection{Hardware and software}
AI ethics play a critical role in the development of cyber and network security, as well as in the design of smart home hardware and software. This section delves deeper into the ethical considerations within smart home systems, focusing on key aspects such as sensors, operating systems, mobile apps, and cloud services.

\subsubsection{Sensors}
With the introduction of low-power sensors, radios, and embedded processors, smart homes are equipped with a vast network of sensors that collaboratively process data to monitor household activities \cite{ding2011sensor}. Sensors like accelerometer-based devices and electronic adherence monitoring systems have transformed health data collection by providing unobtrusive, real-time monitoring of behaviors related to psychological health, such as sleep, physical activity, and medication adherence \cite{psihogios2024ethical}. Given the integration of healthcare within smart homes, ensuring the security and privacy of sensitive data is essential.

Time-critical IoT \cite{taher2021secure} applications in healthcare, for instance, demand secure, real-time access to private information, making authentication in WSNs a priority. A secure, lightweight three-factor authentication (3FA) protocol \cite{taher2021secure} for IoT WSNs, combining hash and XOR \cite{taher2021secure} operations, and integrating biometrics and user password authentication for enhanced security. The protocol's efficiency, security, and functionality were validated through simulations, demonstrating its superiority over existing protocols in terms of communication and computation overhead \cite{taher2021secure}.

WSNs \cite{shin2019lightweight} form the backbone of smart home systems, monitoring environments and controlling devices. However, these networks face significant security threats. An improved lightweight three-factor authentication and key agreement scheme has been proposed \cite{shin2019lightweight} to address weaknesses in earlier schemes, providing detailed security and performance analysis compared to related methods. Additionally, novel privacy-preserving methods \cite{ali2022novel} for smart grid-based Home Area Networks use homomorphic encryption and hash functions to aggregate data and detect attacks, fostering trust through transparent communication.

\subsubsection{Operating Systems}
Smart home IoT devices, often embedded with microcontroller unit (MCU) based systems, rely on specialized operating systems to manage hardware, software, and communication \cite{mocrii2018iot}. AI ethics for IoT operating systems revolve around privacy, security, transparency, accountability, and fairness in the management of smart devices.

\begin{table*}[htbp!]
\centering
\tiny
\caption{Overview of IoT Operating Systems with AI Ethics Considerations, Programming Languages, and Models}
\label{tab:iot_os}
\begin{tabular}{p{1cm}p{2cm}p{1.5cm}p{3.5cm}p{1.8cm}p{2cm}}
\hline
\textbf{Operating System} & \textbf{Key Features} & \textbf{Primary Use Case} & \textbf{AI Ethics Considerations} & \textbf{Programming Language(s)} & \textbf{Programming Model} \\
\hline
Contiki & Low-power, memory-constrained devices, modular design & Research, industrial IoT & Emphasizes secure data transmission with minimal data collection; limited dynamic updates for ethical consistency & C (main), optional C++ & Event-driven with optional multi-threading \\

RIOT & Micro-kernel, POSIX-compliant, highly modular & General IoT applications & Transparent updates enable ethical data management and adaptable privacy; lightweight and secure for diverse devices & C (core), optional C++ for libraries & Micro-kernel with multi-threading support \\

TinyOS & Event-driven, low-power, component-based & Wireless sensor networks (WSN) & Prioritizes privacy with efficient data management and security for low-power networks & nesC (custom C dialect) & Event-driven \\

FreeRTOS & Real-time kernel, supports low-resource devices & Embedded systems (industrial, hobbyist) & Community-driven approach promotes transparency and accountability for ethical use; low power for environmentally conscious designs & C & Real-time, priority-based task scheduling \\

Zephyr & Scalable RTOS, supports multiple architectures & IoT, medical devices & Built-in security and data privacy; strict protocols for ethical data practices, especially in health tech & C (main), optional C++ for application layers & Real-time kernel with multi-threading \\

Mbed OS & High security, IoT connectivity, ARM support & IoT device development & Enforces data privacy via strong security layers; built-in management tools support ethical data usage & C++ (main), some C for low-level tasks & Real-time kernel with thread management \\

NuttX & Standards-compliant, small footprint, real-time capabilities & Standards-compliant embedded systems & Ensures transparency with industry compliance, reliable for ethical data handling and privacy & C (core), optional C++ for apps & Real-time kernel with POSIX multi-threading \\

OpenWrt & Linux-based, highly customizable, networking focus & Networking devices, routers & Privacy and security by default, including firewall features, support ethical networking & C (core), optional C++, shell scripting for custom configurations & Monolithic kernel with multi-process support \\

Android Things & Android-based, user-friendly, secure platform & Consumer electronics, smart devices & Leverages Android’s extensive security infrastructure, ensuring data privacy for consumer applications & Java (main), C++ for performance-critical tasks (NDK) & Monolithic kernel with multi-threading \\
\hline
\end{tabular}
\end{table*}

Table ~\ref{tab:iot_os} outlines key IoT operating systems, highlighting their primary use cases, programming languages, and ethical considerations. Each system—whether Contiki, RIOT, or Android Things—plays a role in ensuring smart homes meet technical and ethical standards, emphasizing privacy, security, transparency, and accountability in IoT management.

\subsubsection{Mobile Apps}
Mobile apps in smart homes, particularly those integrating AI, present ethical challenges surrounding privacy and security. These apps control IoT devices in a Home Area Network, typically relying on Wi-Fi routers for device authentication and protection against unauthorized apps \cite{demetriou2017hanguard}.

HanGuard \cite{demetriou2017hanguard} is a technique that controls communication between IoT devices and their associated apps, employing a software-defined networking (SDN)-like method for fine-grained protection. It requires no changes to IoT devices, apps, or phone operating systems, but ensures that unauthorized access attempts are blocked through router-based verifications. This approach exemplifies how smart home applications can uphold privacy and security without compromising user convenience or system performance.

The integration of AI in smart home mobile apps offers significant benefits, but it also raises critical ethical concerns. Addressing these requires stringent privacy and security measures to ensure that mobile technologies improve daily life without compromising ethical standards.

Incorporating AI ethics into smart home detection technologies is crucial for ensuring fairness, transparency, security, and privacy. Addressing these ethical concerns through advanced algorithms—such as those ensuring algorithmic fairness, diverse datasets, explainable AI, federated learning, adversarial robustness, and secure AI—greatly enhances the reliability and performance of smart home systems. As smart homes evolve, prioritizing cybersecurity, network integrity, and the ethical design of hardware and software will be essential for safeguarding users against potential threats. This comprehensive approach ensures that AI-driven smart home technologies are deployed responsibly, enhancing quality of life while protecting user rights and data integrity.

\section{Challenges}

The adoption of smart home-based detection technologies brings several ethical challenges that require thorough consideration. While these technologies are designed to enhance security, health monitoring, and convenience, they also introduce complex dilemmas surrounding privacy, consent, autonomy, fairness, transparency, and cybersecurity \cite{chung_ethical_2016}. This section summarizes the key challenges based on the reviews in the previous sections, explaining their significance and why many remain unresolved.

\textbf{Privacy Concerns:}
Smart home systems collect vast amounts of personal data, leading to significant privacy concerns. These systems gather sensitive information about users' behaviors, health, and daily routines. Although encryption and local server hosting can mitigate privacy risks by isolating networks from the internet, privacy breaches remain a pressing issue. Unanticipated data leaks or hacking incidents pose persistent threats, highlighting the ongoing need for stronger safeguards. Privacy is critical because unauthorized access to personal data can result in identity theft, financial loss, and a loss of trust between users and technology providers \cite{birchley_smart_2017}.

\textbf{Consent and Autonomy:}
Ensuring informed consent and safeguarding user autonomy are fundamental ethical imperatives in the development of smart home technologies. It is essential for users to clearly understand and voluntarily agree to data collection practices. However, the inherent complexity of smart home systems often hinders users' ability to fully comprehend how their data is being collected, processed, and utilized, thereby limiting their capacity to make truly informed decisions. This gap is critical, as inadequate communication undermines users' ability to exercise their autonomy effectively. Although strides have been made to enhance consent mechanisms, bridging the divide between user comprehension and the sophistication of these technologies remains a pressing challenge \cite{scott2003autonomy}.

\textbf{Algorithmic Fairness and Transparency:}
The use of advanced AI algorithms in smart homes raises concerns about algorithmic fairness and transparency. If not carefully designed, these algorithms may inadvertently perpetuate biases, leading to unequal outcomes for different users. Furthermore, the opacity of AI systems makes it difficult for users to understand how decisions are made, potentially eroding trust in the technology. Addressing fairness, accountability, and transparency in AI systems is essential to ensure that smart home technologies remain equitable and trustworthy. While progress is being made to improve algorithmic transparency, achieving full clarity in these systems remains a substantial challenge \cite{shin2019role}.

\textbf{Cybersecurity Threats:}
Cybersecurity is a critical issue for smart home technologies, as vulnerabilities within these systems could lead to severe personal, financial, or even physical harm. The interconnected nature of smart homes increases the risk of cyberattacks. Despite the implementation of advanced security measures, such as blockchain and secure networks, cybercrime continues to rise, and attackers continuously find new ways to exploit system weaknesses. As a result, cybersecurity is a constantly evolving challenge that requires ongoing innovation to stay ahead of emerging threats. The importance of robust cybersecurity measures lies in protecting personal data, maintaining user trust, preventing potentially devastating security breaches \cite{rajasekharaiah2020cyber}.

Despite significant progress in addressing ethical challenges, many issues remain unresolved due to the complexity of smart home technologies and the rapidly evolving landscape of cyber threats and AI ethics. Persistent concerns surrounding privacy, informed consent, fairness, transparency, and cybersecurity demand ongoing effort and innovation. Addressing these challenges is critical to ensuring that smart homes are not only functional but also ethically aligned, fostering trust and promoting sustainable adoption. The future focus includes: (1) Developing methods and security frameworks to safeguard user data. 
(2) Establishing universal AI ethics standards for smart home platforms. 
(3) Continuous monitoring and auditing to ensure fairness in AI algorithms. 
(4) Examining ethical aspects of innovations like BCIs and XR in smart homes. 
(5) Fostering cooperation among technologists, ethicists, and policymakers for ethical guidelines. 

\section{Conclusions}
The concept of "SMART" has evolved far beyond basic home automation, now encompassing advanced AI-driven technologies designed to enhance convenience, security, energy efficiency, and quality of life. Leveraging AI, sensors, and machine learning, these systems personalize environments by integrating interconnected devices—such as lighting, HVAC, and security systems—that communicate seamlessly via networks like Bluetooth and Wi-Fi. This convergence of technologies has transformed smart homes into adaptive, intelligent ecosystems tailored to user needs.

With the rapid adoption of AI detection technologies, including image, facial, and voice recognition, the ethical dimensions of AI in the IoT ecosystem have come into sharper focus. This paper explores these considerations through the lens of the User Requirements Notation (URN), offering a quantitative literature review that highlights the intersection of AI and ethics. The review identifies key challenges, such as decision-making accountability, bias, transparency, and risk, while proposing guidelines for human-centered design. It outlines fundamental principles—Privacy by Design, Algorithmic Fairness, Transparency, and Accountability—as the cornerstones of ethical AI-driven smart home systems. By presenting a foundational framework, this paper aims to guide researchers and developers in creating ethical, user-centric solutions for the future of smart homes, ensuring technological advancements are balanced with societal values and user trust.

\bibliographystyle{acm}
\bibliography{references}

\end{document}